\documentclass[12pt]{article}%
\usepackage{amssymb}
\usepackage{times}
\usepackage{amsfonts}
  \usepackage{colortbl}
  \usepackage{color}
\usepackage{amsmath}
\usepackage{openbib}
\usepackage[round]{natbib}
\usepackage[nohead]{geometry}
\usepackage[singlespacing]{setspace}
\usepackage[bottom, multiple]{footmisc}
\usepackage{indentfirst}
\usepackage{endnotes}
\usepackage{geometry}
\usepackage{rotating}
\usepackage{booktabs}
\usepackage{multirow}
\usepackage{pifont}
\usepackage{amssymb}
\usepackage{amsmath}
\usepackage{graphicx}
\usepackage{setspace}
\usepackage{rotating}
\usepackage[dvips]{epsfig}
\usepackage{enumerate}
\usepackage{tabularx}
\usepackage{graphicx}%
\usepackage{rotating}
\usepackage[font=small,labelfont=bf]{caption}
\makeatletter
\def\@biblabel#1{\hspace*{-\labelsep}}
\makeatother
\begin{document}

\title{Smooth Nonparametric Bernstein Vine Copulas}
\author{Gregor N.F. Wei\ss \footnote {Address: Corresponding author; Otto-Hahn-Str. 6a, D-44227 Dortmund, Germany, telephone:\ +49 231 755 4608, e-mail:
\textit{gregor.weiss@tu-dortmund.de}}\medskip\\{\normalsize Juniorprofessur f\"{u}r Finance, Technische Universit\"{a}t Dortmund}\medskip\\
Marcus Scheffer \footnote {Address: Otto-Hahn-Str. 6a, D-44221 Dortmund,
Germany, telephone:\ +49 231 755 4231, e-mail:
\textit{marcus.scheffer@tu-dortmund.de}. We are grateful to Janet Gabrysch, Sandra Gabrysch, Daniel Hustert, Felix Irresberger and Janina M\"uhlnickel for their outstanding research assistance. Support by the Collaborative Research Center "Statistical Modeling of Nonlinear Dynamic Processes" (SFB 823) of the German Research Foundation (DFG) is gratefully acknowledged.}\medskip\\{\normalsize Lehrstuhl f\"{u}r Investition und Finanzierung, Technische Universit\"{a}t Dortmund}
}
\maketitle

\sloppy%

\onehalfspacing

\textbf{Abstract}
We propose to use nonparametric Bernstein copulas as bivariate pair-copulas in high-dimensional vine models. The
resulting smooth and nonparametric vine copulas completely obviate the error-prone need for choosing the pair-copulas from parametric copula families. By means of a simulation study and an empirical analysis of financial market data, we show that our proposed smooth nonparametric vine copula model is superior to competing parametric vine models calibrated via Akaike's Information Criterion.
\vspace{1cm}


\strut
\thispagestyle{empty}

\pagebreak%

\onehalfspacing


\section{Introduction}
\setcounter{page}{1}
Following the growing criticism of elliptical models, copulas have emerged both in insurance and risk management as a powerful alternative for modeling the complete dependence structure of a multivariate distribution. Since the seminal work by \cite{embrechts:2002}, the literature on copulas and their use in risk management applications has grown exponentially with several studies concentrating on statistical inference and model selection for copulas \citep[see, e.g.,][]{kim:2007,genestelal:2009} as well as applications \citep[see, e.g.,][]{Chan,grundke:2012,yea}.\footnote{A literature review with a special emphasis on finance-related papers using copulas is given by \cite{genest:2009}. An overview of the different branches of the copula literature is given by Embrechts (2009).}

As simple parametric models are often not flexible enough to model high-dimensional distributions, recent works by \cite{joe:1996, joe:1997}, \cite{bedford:2001,bedford:2002} and \cite{whelan:2004} have proposed copula models which are highly flexible but at the same time still tractable even in higher dimensions. Most notably, vine copulas (also called pair-copula constructions, PCC in short) have emerged as the most promising tool for modeling dependence structures in high dimensions.\footnote{Competing modeling concepts like nested and hierarchical Archimedean copulas are analyzed by \cite{aas:2009b} as well as \cite{fischer:2009}. They conjecture that vine copulas should be preferred over nested or hierarchical Archimedean copulas.} Vine copulas consist of a cascade of conditional bivariate copulas (so called pair-copulas) which can each be chosen from a different parametric copula family. As a result, vine copulas are extremely flexible yet still tractable even in high dimensions as all computations necessary in statistical inference are performed on bivariate data sets \citep[see][for a first discussion of vine copulas in an applied setting]{aas:2009a}.

Similar to the bivariate case,\footnote{See \citet{genestelal:2009}, \citet{kole:2007} and \citet{weiss:2011,weiss:2012rqfa} for discussions of the problem of selecting the best fitting parametric copula.}\ the correct selection of the parametric constituents of the vine, i.e., the pair-copulas, is crucial for the correct specification of a vine copula model. In case of the popular C- and D-vine specifications, the calibration and estimation of a $d$-dimensional vine requires the selection and estimation of $d(d-1)/2$ different pair-copulas from the set of candidate bivariate parametric copula families. Thus, a vine model's increased flexibility only comes at the expense of an increased model risk. 


As a remedy, recent studies have suggested to select the parametric pair-copulas based on graphical data inspection and goodness-of-fit tests \citep{aas:2009a} and to employ sequential heuristics based on Akaike's Information Criterion (AIC) \citep[see, e.g.,][]{brech:2012,dissmann:2011}. \citet{kurowicka:2010} and \citet{brech:2012} propose strategies for simplifying vines by replacing certain pair-copulas by the independence copula (yielding a \textit{truncated} vine copula) or the Gaussian copula (yielding a \textit{simplified vine}). Finally, \citet{hobaekhaff:2012} propose the use of empirical pair-copulas in vine models to circumvent the problem of selecting parametric pair-copulas. 



In this paper, we use the recently proposed nonparametric Bernstein copulas \citep{sancetta:2004,pfeifer:2009,diers:2012}
\ as pair-copulas yielding smooth nonparametric vine copula models that do not require the specification of parametric families. Thus, we extend the ideas laid out by \citet{hobaekhaff:2012} by using an approximation to the empirical pair-copulas. In contrast to their work, however, we approximate the pair-copulas not only nonparametrically but also by the use of continuous functions.\footnote{Using smooth functions to approximate the true underlying dependence structure is in line with our intuition. The superiority of smooth nonparametric approximations of the copula over simple empirical copulas, however, is also found by \citet{shen:2008}. They argue that the improved approximation by the linear B-spline copulas is due to their Lipschitz continuity.}\ In addition, especially the Bernstein copula has recently attracted attention in insurance modeling \citep{diers:2012} and has already proven its merits in an applied setting. Therefore, the contributions of the proposed smooth nonparametric vine copulas are twofold: First, the use of Bernstein 
\ copulas completely obviates the need for the error-prone selection of pair-copulas from pre-specified sets of parametric copulas. The resulting smooth and nonparametric vine copulas do not only constitute extremely flexible tools for modeling high-dimensional dependence structures, they are also characterized by a smaller model risk than their parametric counterparts. Second, Bernstein
\ copulas have been shown to improve on the estimation of the underlying dependence structure by competing nonparametric empirical copulas.\footnote{For example, Bernstein copulas provide a higher rate of consistency than other common nonparametric estimators and do not suffer from boundary bias \citep{kulpa:1999,sancetta:2004,diers:2012}. Similarly, other approximations as, e.g., linear B-spline copulas have also been shown to yield lower average squared approximation errors than competing discrete approximations \citep{shen:2008}.}\ The modeling of a vine model's pair-copulas by the use of smooth approximating functions is thus a natural extension of recently proposed (highly discontinuous) empirical pair-copulas.

The usefulness of the proposed Bernstein
\ vines is demonstrated by means of a simulation study as well as by forecasting and backtesting the Value-at-Risk (VaR) for multivariate portfolios of financial assets. 

The results presented in this study show that our proposed vine copula model with smooth nonparametric Bernstein pair-copulas outperforms the benchmark model with parametric pair-copulas in higher dimensions with respect to the accuracy and numerical stability of the approximation to the true underlying dependence structure. While our nonparametric vine copula model yields worse average squared errors than a benchmark vine copula calibrated by selecting parametric pair-copulas based on AIC values in lower dimensions (e.g., $d=3,5,7$) in our simulations, this result is reversed in higher dimensions. For random vectors of dimension $d=11$ and higher, the parametric benchmark broke down in more than $50\%$ of the simulations due to either the numerical instability of the parameter and AIC estimation or simply due to extremely inaccurate approximations caused by badly selected parametric pair-copulas. In higher dimensions (i.e., the main field of application of vine copulas), our nonparametric modeling approach is thus clearly superior to a parametric vine copula model. Our risk management application, however, shows that even in lower dimensions ($d=5$) our nonparametric model yields VaR forecasts that are not rejected by a range of formal statistical backtests. Consequently, the slightly worse approximation errors of our nonparametric model in lower dimensions do not seem to affect the modeling of a given dependence structure too severely thus underlining the usefulness of our proposed model.

The remainder of this article is structured as follows. Section 2 introduces vine copulas as well as Bernstein copulas which we employ as pair-copulas in the vines. Section 3 presents the results of a simulation study on the approximation errors of both our nonparametric Bernstein vine copula model as well as a heuristically calibrated parametric benchmark model. In Section 4, we conduct an empirical analysis for a five-dimensional financial portfolio. Concluding remarks are given in Section 5.

\section{Smooth nonparametric vine copulas}\label{sec:firstpart}
The purpose of this section is to shortly introduce the fundamentals of vine and Bernstein copulas.

\subsection{Vine copulas}
Vine copulas are obtained by hierarchical cascades of conditional bivariate copulas and are characterized by an increased flexibility for modeling inhomogeneous dependence structures in high dimensions. Here, we only present the basic definition as well as some basic properties of two popular classes of vines, i.e., C- and D-vines,\footnote{The classes of C- and D-vines are subsets of the so-called regular vines \citep[or R-vines in short, see][]{brech:2012}. We do not consider other types of R-vines in this paper but note that our proposed use of smooth Bernstein and B-spline copulas as pair-copulas can also be extended to other subsets of R-vines.} which will be used later on in our application to financial market data. Readers interested in a more rigorous examination of vine copulas and their properties are referred to the excellent studies by \citet{joe:1996,joe:1997} and \citet{bedford:2001,bedford:2002}.

Starting point is the well-known observation that a joint probability density function of dimension $d$ can be decomposed into
\begin{equation}\label{decomposed-density}
f(x_1,\ldots,x_d)=f(x_1)\cdot f(x_2|x_1)\cdot f(x_3|x_1,x_2)\cdot\ldots\cdot f(x_d|x_1,\ldots,x_{d-1}).
\end{equation}
Each factor in this product can then be decomposed further using a conditional copula, e.g.
\begin{equation}\label{decomposed-density2}
f(x_2|x_1)=c_{12}(F_1(x_1),F_2(x_2))\cdot f_2(x_2)
\end{equation}
with $F_i(\cdot)$ being the cumulative distribution function (cdf) of $x_i$ ($i=1,\ldots,d$) and $c_{12}(\cdot)$ being the (in this case unconditional) copula density of $(x_1,x_2)$.

Going further down the initial decomposition, the conditional density $f(x_3|x_1,x_2)$ could be factorised via
\begin{equation}\label{decomposed-conddensity}
f(x_3|x_1,x_2)=c_{23|1}(F_{2|1}(x_2|x_1),F_{3|1}(x_3|x_1))\cdot c_{13}(F_1(x_1),F_3(x_3))\cdot f_3(x_3)
\end{equation}
with $c_{23|1}(\cdot)$ being the conditional copula of $(x_2,x_3)$ given $x_1$.

Finally, substituting the elements of the initial decomposition in \eqref{decomposed-density} with the conditional copulas, we get for dimension $d=3$
\begin{eqnarray}\label{3dvine}
f(x_1,x_2,x_3)&=&c_{23|1}(F_{2|1}(x_2|x_1),F_{3|1}(x_3|x_1))\\
&\cdot& c_{12}(F_1(x_1),F_2(x_2))\notag\\
&\cdot& c_{13}(F_1(x_1),F_3(x_3))\notag\\
&\cdot& f_1(x_1)\cdot f_2(x_2)\cdot f_3(x_3)\notag
\end{eqnarray}
with $c_{12}$, $c_{13}$ and $c_{23|1}$ as \textit{pair-copulas}. Note that as there are several possible decompositions of the conditional distributions, the joint density of $x_1,\ldots,x_d$ can also be represented by different vine decompositions depending on the variables one chooses to condition on. 

\citet{bedford:2001,bedford:2002} propose representing these decompositions of a $d$-dimensional joint density as a nested set of trees where two nodes are joined by an edge in tree $j+1, j=1,\ldots,d-1,$ only if the corresponding edges in tree $j$ share a common node. Consequently, there are $d-1$ trees, where tree $j$ has $d+1-j$ nodes and $d-j$ edges with each edge corresponding to a pair-copula density, i.e., a density of a conditional bivariate parametric copula. 
In a Canonical or C-vine, each tree has a unique node (without loss of generality this is node $1$) that is connected to all other nodes yielding the representation
\begin{equation}\label{decomposed-density c-vine}
f(x_1,\ldots,x_d) = \prod_{k=1}^d f_k(x_k) \prod_{j=1}^{d-1} \prod_{i=1}^{d-j} c_{i,i+j | i+1,\ldots, i+j-1}(F(x_i|x_{i+1}, \ldots, x_{i+j-1}), F(x_{i+j}|x_{i+1}, \ldots, x_{i+j-1}))
\end{equation}
where the subscript $j$ identifies the tree, while $i$ runs over all edges in each tree.
In contrast to this, in a D-vine, no node in any tree $T_j$ is connected to more than two edges yielding the decomposition
\begin{equation}\label{decomposed-density d-vine}
f(x) = \prod_{k=1}^d f_k(x_k) \prod_{j=1}^{d-1} \prod_{i=1}^{d-j} c_{j,j+i | 1,\ldots, j-1}(F(x_j|x_{1}, \ldots, x_{j-1}), F(x_{j+i}|x_{1}, \ldots, x_{j-1})).
\end{equation}

\begin{figure}[htbp]

\begin{center}
\includegraphics[width=15.0cm]{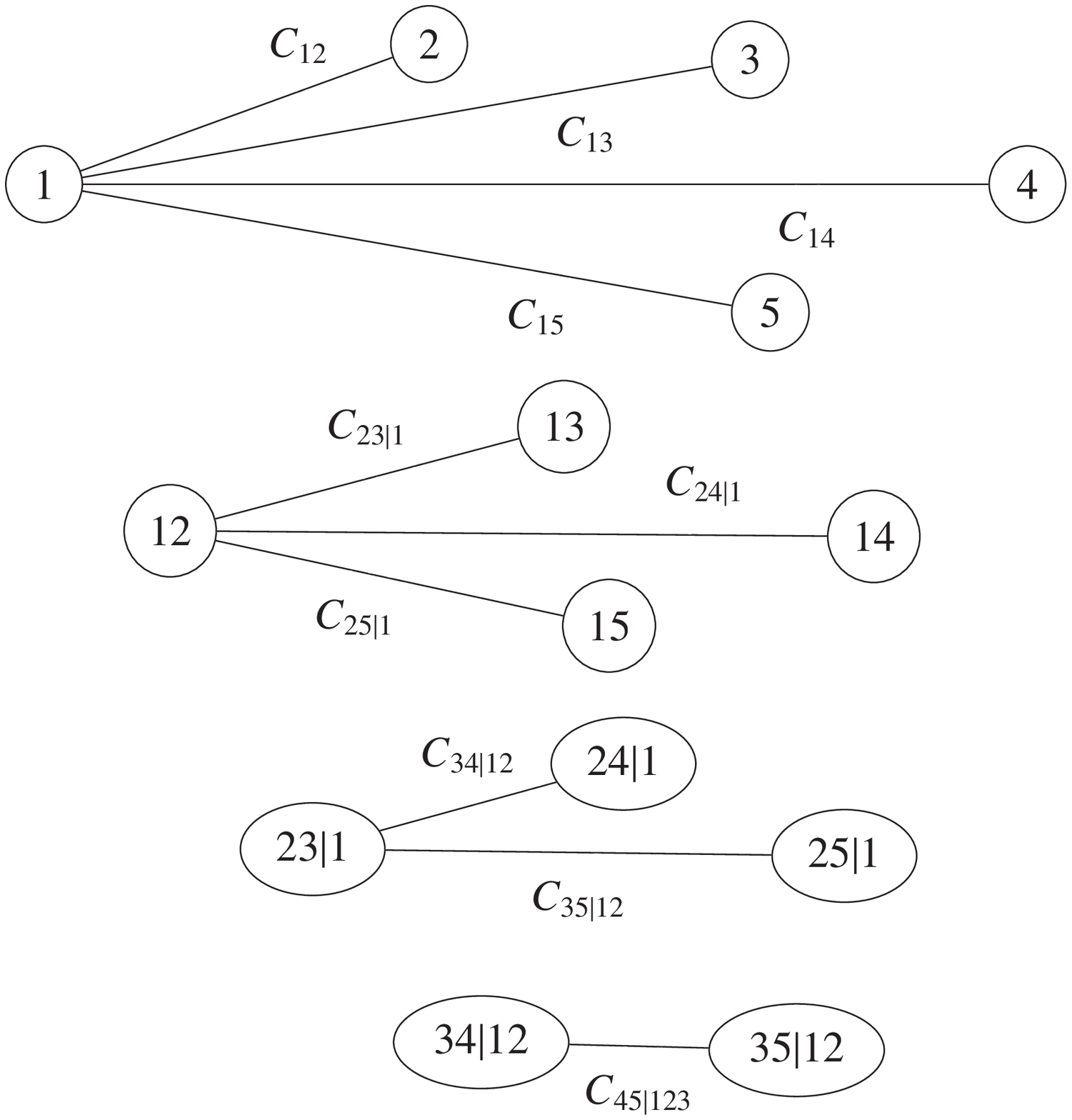}
\end{center}
\caption{{\bf Five-dimensional C-vine copula.} The figure shows an example of a five-dimensional C-vine copula with five random variables, four trees and ten edges. The nodes in the first tree correspond to the five random variables that are being modeled and each edge corresponds to a bivariate conditional or unconditional pair-copula.}
\label{fig:vine1}
\end{figure}

\begin{figure}[htbp]
\begin{center}
\includegraphics[width=15.0cm]{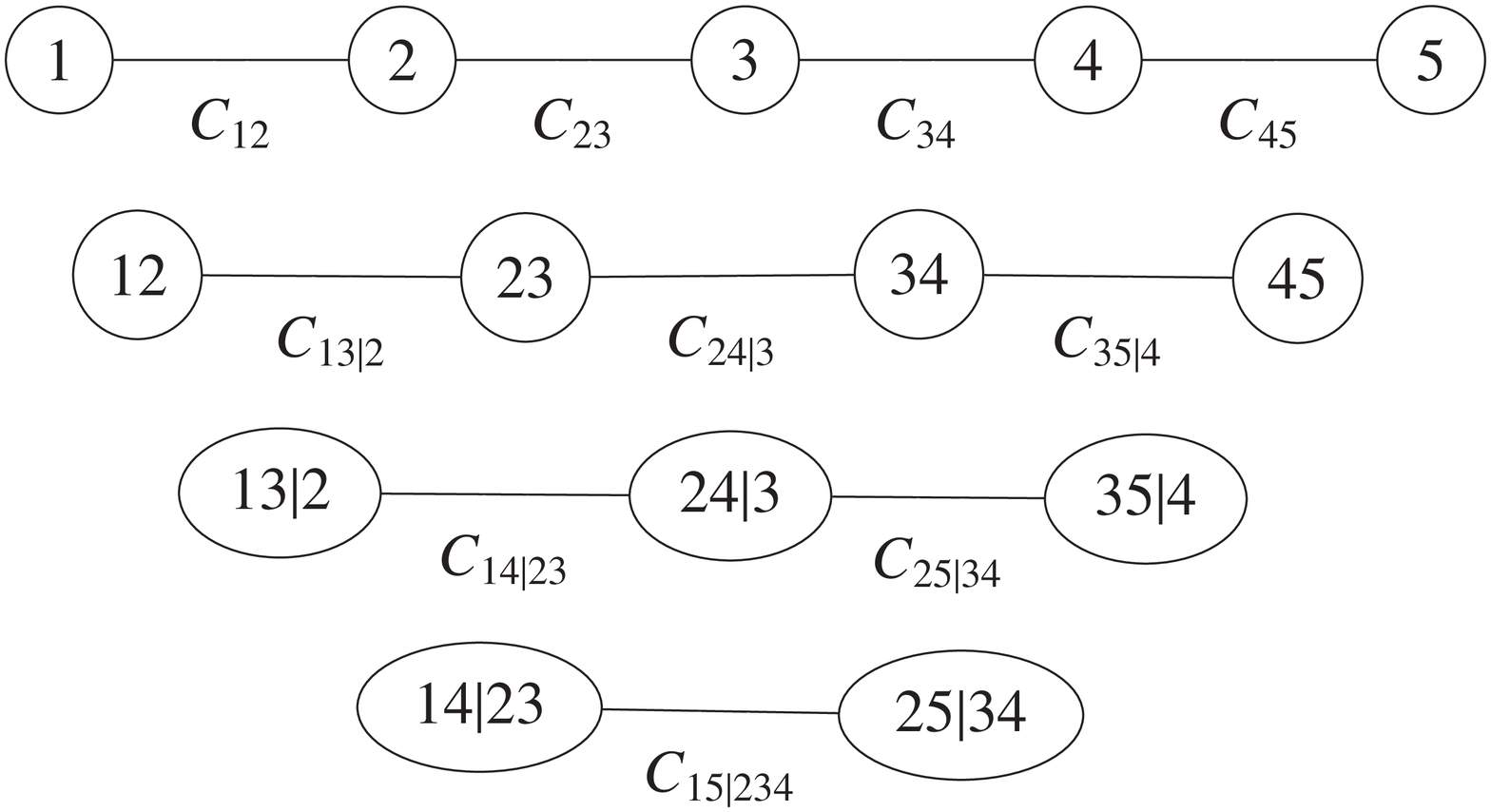}
\end{center}
\caption{{\bf Five-dimensional D-vine copula.} The figure shows an example of a five-dimensional D-vine copula with five random variables, four trees and ten edges. The nodes in the first tree correspond to the five random variables that are being modeled and each edge corresponds to a bivariate conditional or unconditional pair-copula.}
\label{fig:vine2}
\end{figure}

Examples of possible decompositions of a five-dimensional random vector via a C- and D-vine copula are shown in Figure \ref{fig:vine1} and \ref{fig:vine2}, respectively.

\begin{center}
--- insert Figures \ref{fig:vine1} and \ref{fig:vine2} here ---\\
\end{center}

Fitting a vine copula model to a given dataset requires three separate steps. First, one needs to select the tree structure of the vine model. For a C- or D-vine, this amounts to the selection of a permutation of the indices $1,\ldots,d$ of the random variables. As such, for a $d$-dimensional random vector their exist $d!/2$ different C- and D-vines, respectively \citep[see][]{aas:2009a}.\footnote{Choosing the best fitting tree structure manually thus quickly becomes unfeasible in higher dimension \citep[see][]{dissmann:2011}.} Once a permutation has been chosen, the structure of the vine is fully specified. Second, the statistician has to select $d(d-1)/2$ bivariate pair-copulas from candidate copula families. In the last step, the parameters of the pair-copulas have to be estimated. To select the optimal tree structure, \citet{dissmann:2011} propose a heuristic procedure in which the tree structure is chosen via a maximum spanning tree algorithm that maximizes the sum of the absolute empirical Kendall's $\tau$ of all possible variable pairs on a given level of the tree. For the selection of the parametric pair-copulas, \citet{brechczado:2012} and \citet{dissmann:2011} propose a sequential heuristic which selects the best fitting parametric copula family for each pair-copula based on the candidate copulas' AIC values. Although this sequential selection of parametric pair-copulas using AIC does not necessarily yield a globally optimal AIC value for the vine, \citet{brechczado:2012} show that this heuristic yields considerably better results than a selection algorithm based on copula goodness-of-fit tests.\footnote{Results by \citet{weiss:2011} and \citet{grundke:2012} underline this finding. While \citet{weiss:2011} shows that goodness-of-fit tests give only little guidance for choosing copulas when one is interested in forecasting quantiles in the extreme tails of a joint distribution, the empirical results of \citet{grundke:2012} cast additional doubt on the ability of copula goodness-of-fit tests to identify stressed risk dependencies.}

In the following, we concentrate on the problem of selecting the bivariate pair-copulas by substituting them with smooth nonparametric estimates of the underlying (pairwise) dependence structures. As a benchmark to our proposed nonparametric method, we employ the sequential heuristic by \citet{brechczado:2012} and \citet{dissmann:2011}. However, we expect our nonparametric approach to improve on the heuristic benchmark for several reasons. First, the nonparametric modeling of the pair-copulas eliminates completely the model risk of choosing an incorrect parametric family for a given pair-copula. Second, \citet{gronneberg:2012} prove that the use of AIC as a model selection criterion is not correct in case rank-transformed pseudo-observations are used (as is common in almost all applications of copulas in finance). Third, as already hinted at by \citet{dissmann:2011}, the incorrect specification of the parametric pair-copulas in the upper levels of a vine's tree structure can lead to a propagation and amplification of rounding errors causing the heuristic to become numerically unstable.

In the next subsection, we define and discuss Bernstein copulas which we use as smooth nonparametric estimates of the pair-copulas in a vine model.




\subsection{Bernstein copulas}
As a nonparametric candidate for the pair-copulas in \eqref{decomposed-density c-vine} and \eqref{decomposed-density d-vine}, we consider the recently proposed Bernstein copulas. In the following, we briefly state some basic mathematical facts on Bernstein polynomials and Bernstein copulas, respectively. The Bernstein polynomials of degree $m$ are defined as 
\begin{equation}
B(m, k, z) = \binom{m}{k} z^k (1-z)^{m-k}, 
\end{equation}
where $k=0, \ldots, m \in \mathbb{N}$ and $0 \leq z \leq 1$. 

As our focus lies on the nonparametric modeling of pair-copulas in vines, we restrict our analysis in the following to bivariate Bernstein copulas. Let $U=(U_1, U_2)$ denote a discrete bivariate random vector with uniform margins over $T_i:=\{0, 1, \ldots\, m_i\}$ with grid size $m_i \in \mathbb{N}$ and $i=1,2$. In our analysis, we later choose $m_1:=m_2:=m=const$. With 
\begin{equation}
p(k_1, k_2):= P(\displaystyle{\bigcap_{i=1}^2 \{U_i = k_i\}}), \quad (k_1, k_2) \in [0,1]^2
\end{equation}
we can then define the Bernstein copula density as 
\begin{equation}\label{bernstein copula density}
c(u_1, u_2):= \sum_{k_1=0}^{m_1-1} \sum_{k_2=0}^{m_2-1} p(k_1, k_2) \prod_{i=1}^2 m_i B(m_i-1, k_i, u_i), \quad (u_1, u_2) \in [0,1]^2.
\end{equation}
\citet{pfeifer:2009} show by integrating expression (\ref{bernstein copula density}) that the cdf of the Bernstein copula is then given by
\begin{eqnarray}\label{bernstein copula}
C(x_1, x_2) & := & \int_0^{x_2} \int_0^{x_1} c(u_1, u_2) d u_1 d u_2 \\
& = & \sum_{k_1=0}^{m_1} \sum_{k_2=0}^{m_2} P(\bigcap_{i=1}^2 \{U_i < k_i\}) \prod_{i=1}^2 B(m_i, k_i, u_i) 
\end{eqnarray}
for $(x_1, x_2) \in [0,1]^2$. Note that in order to smoothly approximate the distribution or density of a copula in (\ref{bernstein copula density}) and (\ref{bernstein copula}), very high degrees for the Bernstein polynomials have to be chosen.

The Bernstein copula as defined above can then be used to approximate the empirical copula process as defined, e.g., by \citet{deheuvels:1979}. To be precise, we approximate nonparametrically the joint distribution of $U=(U_1, U_2)$ by using a bivariate sample $\{(X_i, Y_i)\}^n_{i=1}$ of the underlying copula of size $n$.
Let $X_{(k)}$ be the $k$th order statistic of the sample. The empirical copula of \citet{deheuvels:1979,deheuvels:1981} is then defined as
\begin{equation}\label{eq:empcop}
C_n (x, y) = \begin{cases}
  C_n \left(\frac{i-1}{n}, \frac{j-1}{n}\right),  & \frac{i-1}{n} \le x < \frac{i}{n}, \frac{j-1}{n} \le y < \frac{j}{n}\\
  1, & x=y=1,
\end{cases}
\end{equation}
where $C_n(x,y)=\frac{1}{n} \sum_{k=1}^n 1_{\{ X_j \le X_{(i)},  Y_j \le Y_{(i)} \}}$, $C_n(0,\frac{i}{n}) = C_n (\frac{j}{n}, 0) = 0$ and $C_n (0, \frac{i}{n}) = 0$ ($i,j=1,2,...,n$). 


To fit the Bernstein copula to the sample from the empirical copula, we first need to calculate the relative frequency of the observations in each target cell of a grid with given grid size $m$. The outcome of this is the contingency table $[a_{kl}]_{k,l=1,\ldots,m}$. Note, however, that the resulting marginals of the data $[a_{kl}]$ do not need to be uniformly distributed and that the resulting approximation could therefore not be a copula. To circumvent this problem, \citet{pfeifer:2009} propose to transform the contingency table $[a_{kl}]$ into a (possibly suboptimal) new contingency table $[x_{kl}]$ with uniform marginals via a Lagrange optimization approach yielding
\begin{equation}
x_{ij} = a_{ij} - \frac{a_{\cdot j}}{m} - \frac{a_{i \cdot}}{m} + \frac{2}{m^2} \qquad \text{for } i,j = 1,\ldots,m,
\end{equation}
where the index $\cdot$ denotes summation. Note that the quality of the Lagrange solution is reduced by an increasing number of the sample size $n$.\footnote{In unreported results, the optimization strategy of \citet{pfeifer:2009} proved to yield only suboptimal results.}\ We therefore chose to employ a different optimization strategy to correct for the non-uniform distribution of the marginals.

Consequently, we calculate the approximation $[x_{kl}]$ to the contingency table $[a_{kl}]$ by solving the following optimization problem:
\begin{eqnarray}\label{opt problem}
& \displaystyle{\sum_{k=1}^m \sum_{l=1}^m} (x_{kl} - a_{kl})^2 \longrightarrow \text{min}\\
\text{subject to} & & \notag\\
& \displaystyle{\sum_{k=1}^m x_{kj} = \sum_{l=1}^m x_{il}} = \frac{1}{m} \qquad \text{and} \qquad x_{ij} \geq 0 \quad \text{for } i,j=1,\ldots,m. 
\end{eqnarray}

To solve for the $[x_{kl}]$, we make use of the quadratic optimization algorithm of \citet{goldfarb:1982,goldfarb:1983}. In preliminary tests, the found solutions to this optimization problem yielded significantly lower quadratic errors than the procedure initially proposed by \citet{pfeifer:2009} thus confirming the need for a more refined optimization strategy.

To use Bernstein copulas as pair-copulas both in our simulation study and the empirical application, we require efficient algorithms for simulating and evaluating the density and distribution of a given vine copula. To this end, we adapt the algorithms initially proposed by \citet{aas:2009a} by substituting the parametric h-hunctions (i.e., the partial derivatives of the copula densities) in these algorithms by the partial derivatives of the fitted bivariate Bernstein copulas.

\section{Simulations}
In this section, we illustrate the superiority of the smooth nonparametric vine model over the sequential heuristic of \citet{brech:2012} and \citet{dissmann:2011} for selecting the pair-copulas in a vine parametrically. In particular, we are interested in the error of the approximations to a pre-specified true copula using both our smooth nonparametric model as well as a vine model calibrated with parametric pair-copulas. The setup of our simulation study follows the procedure laid out in \citet{shen:2008}, but differs in that way that we also consider a (heuristically calibrated) parametric benchmark approximation to the true vine copula. 

As a measure for the approximation error, we compare the pre-specified true pair-copulas of the vine model with the parametric and nonparametric approximations and use the average squared error (ASE) of the cdfs of all bivariate pair-copulas each taken at $m_1\times m_2$ uniform grid points in $I^2$, i.e.,
\begin{equation}
\label{ASE}
ASE:=\frac{2}{d(d-1)}\frac{1}{m_1\cdot m_2}\sum_{i=1}^{d(d-1)/2}\sum_{j=1}^{m_1}\sum_{k=1}^{m_2} \left( \hat{C}_i\left(\frac{j}{m_1+1},\frac{k}{m_2+1}\right)-C_i\left(\frac{j}{m_1+1},\frac{k}{m_2+1}\right)    \right)^2
\end{equation}
where $C_i$ is the cdf of a pre-specified pair-copula from which we simulate a random sample of size $n$ and $\hat{C}_i$ is a (parametric or nonparametric) approximation to the pair-copula $C_i$ computed on the basis of the simulated random sample.

In the simulation study, we consider two sample sizes $n=200$ and $n=500$ to assess the decreasing effect of the sample size on the approximation error. Furthermore, we analyze the effect of the type (C- or D-vine) as well as the dimensionality of the vine model on the approximation errors. To be precise, we simulate random samples from vines of dimension $d=3,5,6,7,11,13$. As the dimension of the vine model increases, so does the number of variables one has to condition on in the pair-copulas of the vine's lower trees. The pair-copulas in the lower trees of the vine, however, are generally more complex to model so that the accurate approximation of the pair-copulas on all levels of the vine constitutes a considerable challenge to our nonparametric approximation.\footnote{This is one reason why \citet{aas:2009a}, \citet{brech:2012} and \citet{dissmann:2011} propose to capture as much dependence of the joint distribution that is to be modeled in the first trees of a vine model. If these pair-copulas are modeled accurately, the remaining pair-copulas in the lower trees can then be truncated or simplified. Furthermore, the truncation and simplification of a vine on the lower levels of the vine's tree limits the potential propagation of rounding errors.}\ At the same time, the curse of dimensionality could additionally complicate the approximation of the pair-copulas thus making the comparison of our approximation for different dimensions a sensible exercise. Finally, we expect the propagation and amplification of rounding errors to increase in higher dimensions possibly leading to the numerical instability of the parametric heuristic.

As candidate parametric copula families from which the pair-copulas of the true vine models are chosen, we use the Gaussian, Student's t, Clayton, Gumbel, Survival Clayton, Survival Gumbel, the rotated Clayton copula (90 degrees) and the rotated Gumbel copula (90 degrees). The pair-copulas as well as their respective parameters in each simulation are chosen randomly. For each sample size, dimension and vine type, we simulate $1,000$ random samples and approximate the data with a vine copula using Bernstein copulas as pair-copulas. As we are only interested in measuring the accuracy of the approximation of the pair-copulas, we calibrate our nonparametric vine using the correct vine type as well as the correct tree structure. As a benchmark, we calibrate a second vine copula by using the sequential heuristic proposed by \citet{brechczado:2012} and \citet{dissmann:2011}. Furthermore, we also compute the fraction of time the sequential procedure breaks down due to either the numerical instability of the evaluation of the likelihood function and the parameter estimation or due to the ASE tending to infinity. The results of the simulations are presented in Table \ref{tab:simu}.

\begin{center}
--- insert Table \ref{tab:simu} here ---\\
\end{center}

The results shown in Table \ref{tab:simu} present several interesting insights into the finite sample properties of both the heuristically calibrated parametric and our proposed nonparametric vine copula models. First, we can see from Table \ref{tab:simu} that for lower dimensions (e.g., $d=3$ and $d=5$) the ASE of our nonparametric is considerably larger than for the parametric model calibrated by sequentially selecting the pair-copulas based on AIC values. With increasing dimension of the random vector, however, we can observe that the approximation error of the parametric model increases disproportionately compared to our proposed nonparametric model. Furthermore, the nonparametric model appears to be able to match the approximation error of the parametric approach for dimensions $d=13$ and higher. Most importantly, the parametric modeling approach becomes highly numerically unstable in higher dimensions. At the same time, our proposed nonparametric vine with Bernstein pair-copulas is extremely reliable yielding acceptable approximations to the true underlying dependence structure even for high-dimensional random vectors. The parametric approach, on the other hand, breaks down in approximately $50\%$ of all simulations for dimension $d=13$ and higher. In many of these cases, the bad approximation (or numerical instability) of the parametric approach was caused by the wrong selection of several parametric families for the pair-copulas in the vine model. Concerning the type of the vine copula model, we find no significant differences between the average approximation errors of the C- or D-vines. As expected, we also find the average approximation error of both the parametric and nonparametric model to be decreasing in the sample size used for estimating both models.

\begin{figure}[htbp]

\begin{center}
\includegraphics[width=8cm]{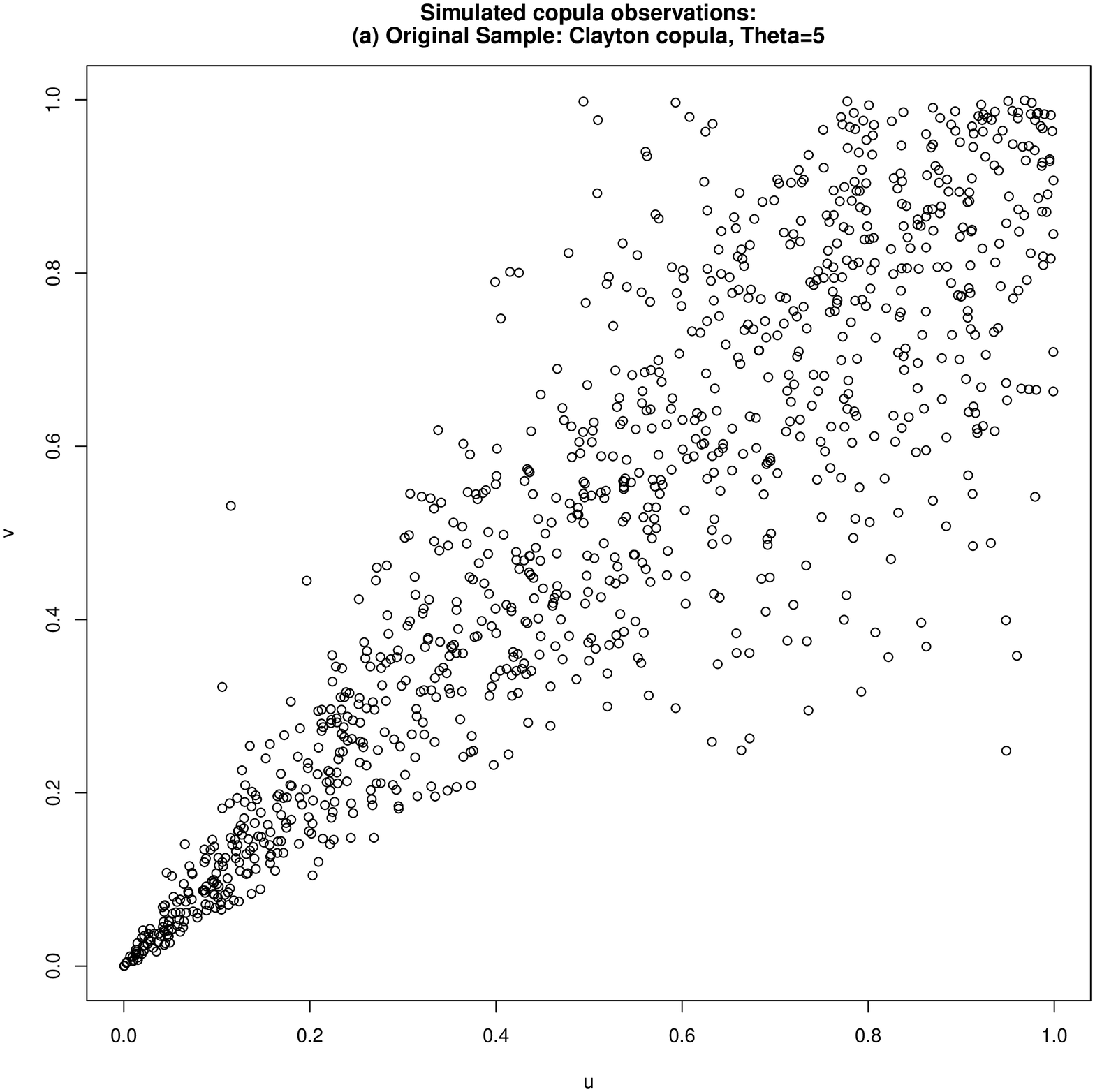}
\includegraphics[width=8cm]{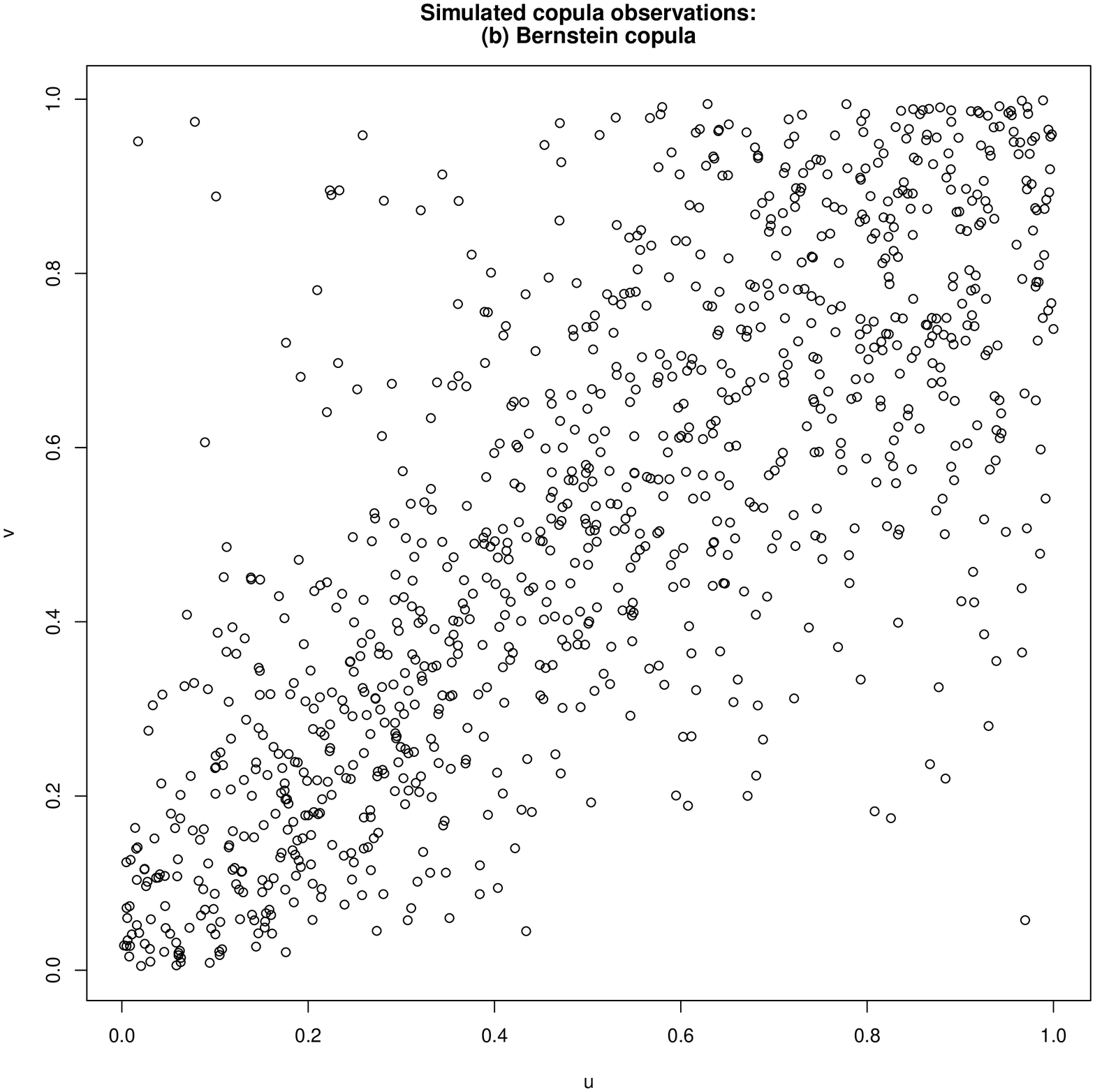}\\
\includegraphics[width=8cm]{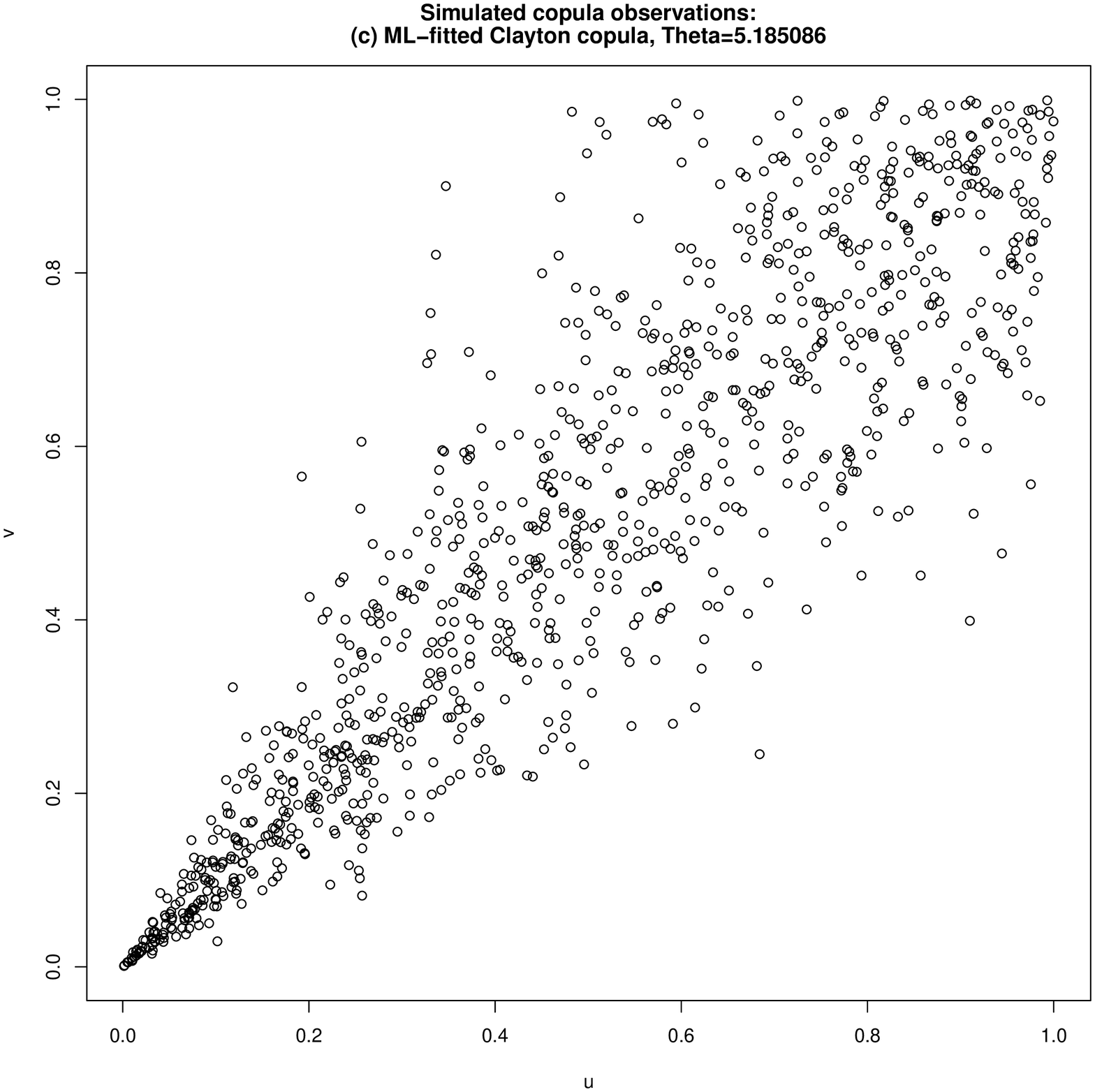}
\includegraphics[width=8cm]{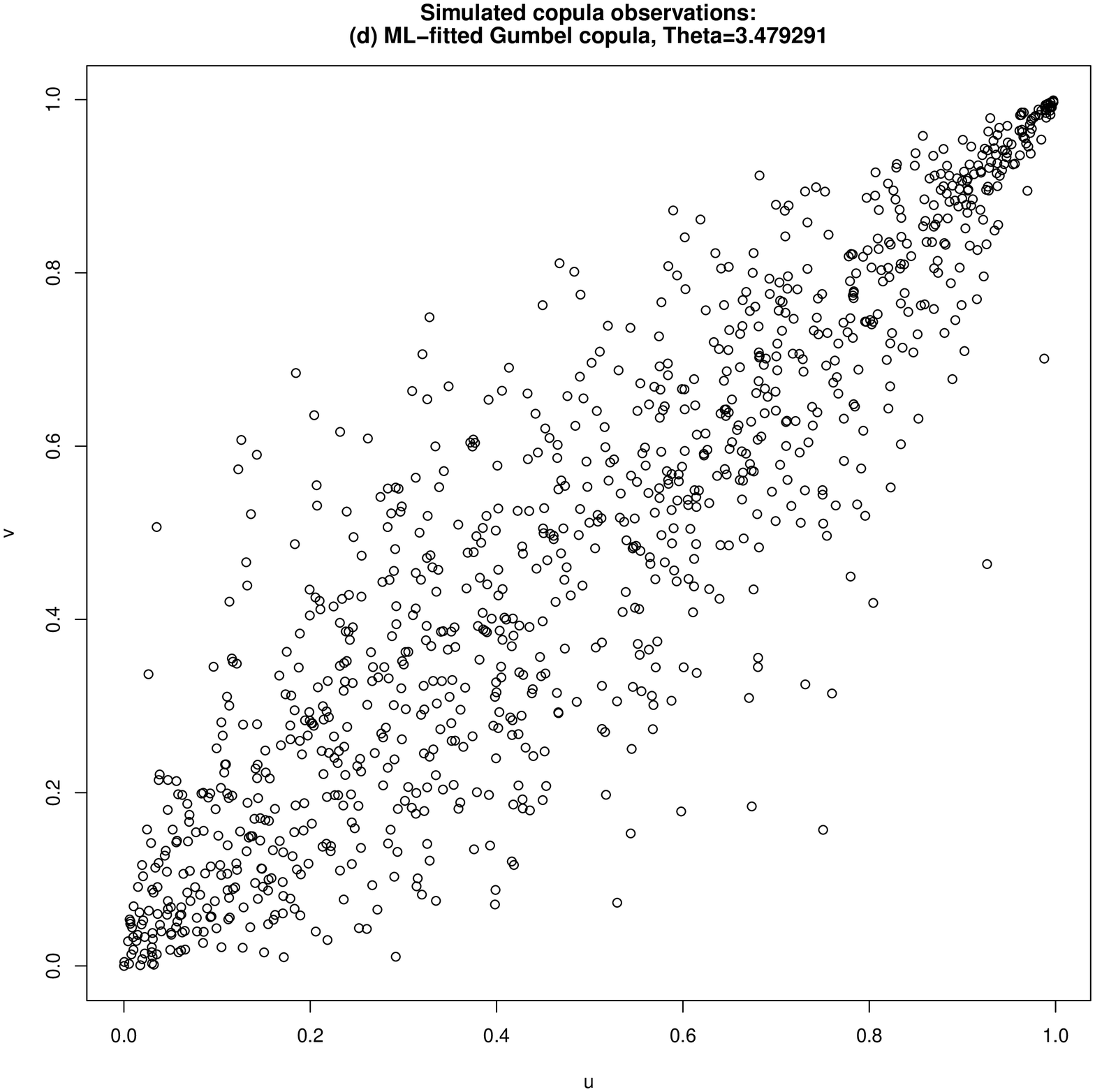}

\end{center}
\caption{{\bf Simulated observations from parametrically and nonparametrically fitted copulas.} The figure shows plots of $n=1000$ observations simulated from different parametric and nonparametric copulas. Panel (a) shows the plot of the original observations simulated from a Clayton copula with parameter $\theta=5$ which are used to calibrate a nonparametric Bernstein copula and two parametric Clayton and Gumbel copulas. Panel (b) shows a sample of simulated observations from a Bernstein copula which was calibrated based on the sample in Panel (a). Panels (c) and (d) show similar plots of simulated samples from a parametric Clayton and Gumbel copula which were fitted via Maximum-Likelihood using the original sample shown in Panel (a).}
\label{fig:approx}
\end{figure}

To further illustrate the finding that the nonparametric model improves on the accuracy of a parametric vine especially in higher dimensions, we plot simulated samples from several parametrically and nonparametrically fitted copulas in Figure \ref{fig:approx} where we assume that the true underlying dependence structure is given by a Clayton copula with parameter $\theta=5$. From this copula, we simulate a random sample of size $n=500$ and fit both a nonparametric Bernstein copula as well as a parametric Clayton and Gumbel copula via Maximum-Likelihood to the data. From all three fitted copulas, we again simulate a random sample and compare the plots of the simulated observations with the original sample.

\begin{center}
--- insert Figure \ref{fig:approx} here ---\\
\end{center}

The plots in Panels (a) and (b) in Figure \ref{fig:approx} show how the Bernstein copula approximately captures the lower tail dependence of the original sample shown in Panel (a). The plot of the Bernstein copula in Panel (b) also shows, however, that the nonparametric approximation of the data sample coincides with a loss in information on the tail behaviour of the true underlying dependence structure. At the same time, Panel (c) underlines the notion that the nonparametric model is not superior to a parametric model in which the parametric copula family has been chosen correctly. If, however, the parametric copula familiy is chosen incorrectly like it is shown in Panel (d), the wrong selection of the parametric copula family can cause considerable approximation errors and a severely inaccurate modeling of the underlying tail dependence. Given no prior information on the parametric copula familiy, the nonparametric Bernstein copula model clearly improves on the fit of an inaccurately fitted parametric model. The nonparametric modeling of the pair-copulas thus seems to be a sensible approach especially when the number of pair-copulas that need to be selected from candidate parametric copula families increases (i.e., with increasing dimension).

\section{Empirical study}
\subsection{Methodology}
\label{sec:meth}
The purpose of our empirical study is to investigate the superiority of the proposed smooth nonparametric vine copula models over the competing calibration strategy based on a sequential selection of parametric pair-copulas via AIC with regard to the accurate forecasting of a portfolio's VaR. The simulation study presented in the previous section has highlighted the finding that our vine model with smooth nonparametric pair-copulas is especially well-suited for dependence modeling in higher dimensions as the selection of parametric pair-copulas becomes numerically unstable due to error propagation and amplification. For low-dimensional problems, the heuristic selection of parametric copulas, however, seems to outperform our nonparametric approach with respect to the ASE of the vine copula's approximation. To show that our nonparametric model matches the results of the parametric heuristic even for low-dimensional problems, we concentrate in our empirical analysis on the VaR forecasts of a five-dimensional portfolio.

Financial data are usually characterized by the presence of both conditional heteroscedasticity and asymmetric dependence in the log returns on financial returns. Therefore, we follow the vast majority of studies on copula models for VaR-estimation \citep{jondeau:2006,fantazzini:2009b,ausin:2010,hafner:2010} and employ standard GARCH(1,1)-models with Student's t-distributed innovations to model the marginal behaviour of our data. Although different specifications of the GARCH model are also possible, results found by \citet{hansen:2005} suggest that the choice of the order of a GARCH model is only of little importance for the model's forecasting accuracy.

Throughout the empirical study, we consider continuous log returns on financial assets with prices $P_t$ $(t=0,1,\ldots,T)$. The assets' log returns $R_t$ are defined by $R_t:=\log(P_t/P_{t-1})$ for $t\geq 1$. Our focus lies on modeling the joint distribution of the $d$ assets, i.e., the joint distribution of the returns $R_{t1},\ldots,R_{td}$.

The marginal behaviour of the assets is modeled by the use of GARCH(1,1)-models with t-distributed innovations. The marginal model is then given by
\begin{eqnarray}
R_{tj}&=&\mu_j+\sigma_{tj}Z_{tj},\\
\sigma^2_{tj}&=&\alpha_{0j}+\alpha_{1j}R^2_{t-1,j}+\beta_j\sigma^2_{t-1,j},\ j=1,\ldots,d;\ t=1,\ldots,T,
\end{eqnarray}
with independent and identically t-distributed innovations $Z_{tj}$. The dependence structure between the $d$ assets is introduced into the model by assuming the vector ${\bf Z}_t=(Z_{t1},\ldots,Z_{td})$ ($t=1,\ldots,T$) of the innovations to be jointly distibuted under a $d$-dimensional copula $C$ with
\begin{equation}
F_{\bf Z}({\bf z};\boldsymbol{\nu}_1,\ldots,\boldsymbol{\nu}_d,\boldsymbol{\omega}|\mathcal{F}_{t-1})=C\left[F_1(z_1;\boldsymbol{\nu}_1|\mathcal{F}_{t-1}),\ldots,F_d(z_d;\boldsymbol{\nu}_d|\mathcal{F}_{t-1});\boldsymbol{\omega}\right]
\end{equation}
where $\boldsymbol{\nu}_1,\ldots,\boldsymbol{\nu}_d$ are the parameter vectors of the innovations, $C$ is a copula and $\boldsymbol{\omega}$ is a vector of copula parameters (in case of the parametric model, otherwise $\boldsymbol{\omega}$ is simply empty).

The parameters of the univariate GARCH-models are estimated via Quasi-Maximum Likelihood Estimation. For the estimation of both the nonparametric vine model as well as the parametric model calibrated by using the pair-copulas' AIC values, we make use of rank-transformed pseudo-observations rather than the original sample as input data.\footnote{For a comparative study on the finite sample properties of different ML-based estimators for copulas, see \citet{kim:2007}. The authors show that absent any information on the true distribution of the marginals, statistical inferences should be based on rank-transformed pseudo-observations.}\ As the main results for copulas only hold for i.i.d. samples, we use the parameter estimates for the univariate GARCH models and transform the original observations into standardized residuals to yield (approximately) i.i.d. observations before computing the pseudo-observations \citep{dias:2009}. 

In our empirical application, we consider an equally-weighted five-dimensional portfolio with returns $R_{p,t}=d^{-1}\sum_{j=1}^dR_{tj}$. The results from our simulation study underline the finding that our proposed vine copula model with Bernstein pair-copulas, on average, yields a better approximation to the empirical copula than the heuristically calibrated parametric model especially in higher dimensions. However, the parametric copula vine model could still outperform our proposed model w.r.t. the forecasting accuracy in low dimensions. In our empirical application, we therefore restrict our analysis to a portfolio consisting of five assets to additionally illustrate the nonparametric Bernstein vine copula model's superiority for low-dimensional problems.

To forecast the portfolio returns, we employ the algorithm presented in the study by \citet{nikoloulopoulos:2011} initially proposed for in-sample forecasting which was extended to out-of-sample forecasting by \citet{weiss:2012rqfa}. 

The aim of the algorithm is the computation of a one-day-ahead forecast for the portfolio return $R_{p,t}$ via Monte Carlo simulation. In a first step, $K=10,000$ observations $u_{T+1,1}^{(k)},\ldots,u_{T+1,d}^{(k)}$ ($k=1,\ldots,K$) from the fitted (parametric or nonparametric) vine copula are simulated. Using the quantile function of the fitted marginal Student's t distributions, the simulated vine copula observations are then transformed into observations $z_{T+1,j}^{(k)}$ from the joint distribution of the innovations. In the next step, the simulated innovations are transformed into simulated returns $R_{T+1,j}^{(k)}=\hat{\mu}_j+\hat{\sigma}_{T+1,j}z_{T+1,j}^{(k)}$ where $\hat{\sigma}_{T+1,j}$ and $\hat{\mu}_j$ are the forecasted conditional volatility and mean values from the previously fitted marginal GARCH models. The MC-simulated forecasts of the portfolio return is then simply given by $R_{T+1,p}^{(k)}=d^{-1}\sum_{j=1}^dR_{T+1,j}^{(k)}$. Sorting the simulated portfolio returns for a given day in the forecasting period and taking the empirical one-day $\alpha$ percentile then yields the forecasted $\alpha\%$-VaR.



To backtest the results of our forecasting, we employ the test of conditional coverage proposed by \citet{chris1} as well as two duration-based tests discussed in \citet{chris2}. \footnote{See \citet{Berk} for an excellent review of different methods for backtesting Value-at-Risk forecasts. A comparison of different backtests can be found in the recent study by \citet{escanciano}.}

All three backtests are based on the hit sequence of VaR-exceedances which is defined by
\[
h_{t,\alpha}:=\left\{
\begin{array}{ll} 1, & \mbox{if $R_{p,t}<$VaR$_{\alpha}(R_{p,t})|\mathcal{F}_{t-1}$} \\
         0, & \mbox{otherwise}
\end{array}\right. .
\]
with $t$ being the time subscript and $\mathcal{F}_{t-1}$ being the set of available information. The test of conditional coverage by \citet{chris1} and \citet{chris2} jointly tests for the correct number of VaR-exceedances (unconditional coverage) and the serial independence of the violations over the complete out-of-sample (independence).\footnote{The test of unconditional coverage has been implicitly incorporated in the Basel Accord for determining capital requirements for market risks, see \citet{Basel}. Consequently, it has since become an industry standard in market risk management, see, e.g., \citet{escanciano}.} Under the null hypothesis of a correct number of VaR-exceedances that are independent over time, the hit sequence is simply distributed as \citep{chris2}
\[
h_{t,\alpha} \sim i.i.d.\ Bernoulli(\alpha).
\]
Then, let $P$ be the length of the out-of-sample, $P_1$ be the number of VaR-exceedances and $P_0$ be the number of days on which the daily VaR forecast was not exceeded, respectively (and consequently $P=P_1+P_0$). The likelihood function for the i.i.d. $Bernoulli$ hit sequence with unknown parameter $\pi_1$ is
\begin{equation}
L(h_{t,\alpha},\pi_1)=\pi_1^{P_1}(1-\pi_1)^{P-P_1}
\end{equation}
and the Maximum-Likelihood estimate of $\pi_1$ is simply given by $\hat{\pi}_1=P_1/P$. The test of unconditional coverage is then given by a likelihood ratio test based on
\begin{equation}
LR_{UC}= -2 \left(\ln L(h_{t,\alpha},\hat{\pi}_1) - \ln L (h_{t,\alpha},\alpha)\right).
\end{equation} 
To test the hypothesis of independently distributed hits, the hit sequence is assumed to follow a first order Markov sequence
with switching probability matrix
\begin{equation}
\Pi = 
\begin{bmatrix}
 1- \pi _{01} & \pi _{01}\\
 1- \pi _{11} & \pi _{11}
\end{bmatrix}
\end{equation}
with $\pi_{ij}$ being the probability of an $i$ on day $t-1$ being followed by a $j$ on the next day $t$ and $i,j\in\left\{1;0\right\}$. Using the likelihood function
\[
L (h_{t,\alpha},\pi_{01},\pi_{11}) = (1 - \pi_{01})^{P_0-P_{01}} \pi_{01}^{P_{01}} (1-\pi_{11})^{P_1-P_{11}}\pi_{11}^{P_{11}}.
\]
where $P_{ij}$ is the number of observations in $h_{t,\alpha}$ where a $j$ follows an $i$ and $i,j\in\left\{1;0\right\}$ and the ML-estimates $\hat{\pi}_{01}=P_{01}/P_0$ and $\hat{\pi}_{11}=P_{11}/P_1$, the likelihood ratio test of the independence of hits is given by
\begin{equation}
LR_{ind}= 2 \left(\ln L(h_{t,\alpha},\hat{\pi}_{01},\hat{\pi}_{11}) - \ln L (h_{t,\alpha},\hat{\pi}_{1})\right).
\end{equation} 
Both tests are then combined via $LR_{CC}=LR_{UC}+LR_{ind}$ to yield the test of conditional coverage. We note here that we do not rely on the asymptotic Chi-squared distribution of the test statistic which is used, e.g., in the study by \citet{hsu:2011}. Although easy to implement, p-values derived under the assumption of the test statistic following a Chi-squared distribution are usually incorrect due to the generally low sample sizes when using hit sequences. Instead, we follow \citet{chris2} in generating approximate p-values via Monte Carlo-simulation.

As an alternative to the test of conditional coverage, \citet{chris2} propose backtests based on the durations between VaR-exceedances. Then, let
\begin{equation}
D_i=t_i-t_{i-1}
\end{equation}
be the duration of time (in trading days) between two subsequent VaR-exceedances where $t_i$ is the time of the $i$th VaR-exceedance. Under the null hypothesis of a correctly specified VaR model, we would expect the process of no-hit durations to have no memory and mean $1/\alpha$. Consequently, the process $D$ of durations should follow an exponential distribution with $f_{exp}(D;\alpha)=\alpha\exp\left(-\alpha D\right)$.\footnote{See \citet{chris2} for details of the backtest and the motivation for using a continuous distribution for the discrete process $D$.} As an alternative hypothesis, \citet{chris2} propose to use the Weibull distribution for the process $D$ with $f_{W}(D;a,b)=a^bbD^{b-1}\exp\left(-(aD)^b\right)$ which nests the exponential distribution from the null hypothesis for $b=1$.

Although this test potentially captures higher order dependence in the hit sequence $h_{t,\alpha}$, the information from the temporal ordering of the no-hit durations is not exploited in the backtest. As a remedy, \citet{chris2} propose a conditional duration-based test based on the Exponential Autoregressive Conditional Duration (EACD) model of \citet{engle:1998}. In the standard EACD (1,0) model, the conditional expected duration $\mathbb{E}_{i-1}(D_i)$ is assumed to follow the process
\begin{equation}
\mathbb{E}_{i-1}(D_i)\equiv\psi_i=\omega+\beta D_{i-1}.
\end{equation}
Again assuming an underlying exponential density with mean equal to one in the null hypothesis, the conditional distribution of the duration is is given by
\begin{equation}
f_{EACD}\left(D_i|\psi_i\right)=\frac{1}{\psi_i}\exp\left(-\frac{D_i}{\psi_i}\right).
\end{equation}
The null hypothesis of independent no-hit durations is then given by $H_0:\beta=0$.

\subsection{Data}

We consider a five-dimensional equal-weighted portfolio consisting of the returns on the EURO STOXX 50 Price Index, 30-year US Treasury Bonds, France Benchmark 10-year Government Bond Index, Gold Bullion LBM and one-month forward Crude Oil Brent. We obtain the data from the \textit{Thomson Reuters Datastream} database. We follow the screening procedure proposed by \citet{inceporter:2006} to control for known sources of data errors in \textit{Datastream}. To be precise, we check whether our data include prices below \$ 1 (which could lead to erroneous log returns due to \textit{Datastream}'s practice of rounding prices) as well as log returns above 300\% that are reversed within one month. Our five univariate time series do not suffer from any of these data errors.

Our sample covers a period of $800$ trading days ranging from June 6, 2009 to July 19, 2012 and thus includes the aftermath of the default of Lehman Bros. as well as the onset of the Sovereign Debt Crisis. We use rolling windows with a length of $500$ trading days for forecasting the one-day-ahead VaR on the following trading day. Our full out-of-sample spans a period of $300$ trading days. Time series plots of the five portfolio constituents as well as the returns on the equal-weighted portfolio are shown in Figure \ref{fig:Hist1}. Panels (a) through (e) show the time series plots of the univariate returns, while Panel (f) shows the time series plot of the portfolio. The initial in-sample is shaded in grey to highlight the out-of-sample consisting of $300$ trading days.
\begin{center}
--- insert Figure \ref{fig:Hist1} here ---\\
\end{center}

The plots in Figure \ref{fig:Hist1} show several distinct features that can complicate VaR-forecasting. First, all plots exhibit the common stylized fact of volatility clusters, e.g., in Panels (a) and (c). Second, overall volatility of the univariate returns differs significantly across our five portfolio constituents. For example, while the returns on the 30-year US treasury bonds are quite volatile in the in-sample and seem to calm in the out-sample, the opposite is true for the France Benchmark 10-year Government Bond Index which exhibits low volatility in the in-sample and a pronounced cluster of high volatility and extreme spikes around November 2011. This last result is, however, not surprising considering the fact that the Sovereign Debt Crisis experienced a climax at that time with the resignation of the Greek and Italian Prime Ministers, early elections in Spain and the expansion of the European Financial Stabilisation Mechanism (EFSM). Similarly, the price of Gold bullion became more volatile in the out-of-sample as well. The plot in Panel (f) shows that the combination of the five individual assets produces a portfolio which exhibits several phases of both high and low volatility as well as sudden extreme spikes in the portfolio's log returns.

\subsection{Results}
Following the methodology presented in Section \ref{sec:meth}, we compute the one-step-ahead forecasts of the portfolio-VaR for each day in the out-of-sample using rolling windows of $500$ trading days. To analyze the differential effect of different confidence levels for the VaR on our models' forecasting accuracy, we forecast the $2\%$-, $5\%$- and $10\%$-VaR for a long and the $97.5\%$-VaR for a short position in the portfolio. Thus, we would expect $6$, $15$, $30$ and $8$ exceedances below the forecasted VaRs, respectively.\footnote{For the short position, VaR-exceedances are defined as returns above the daily VaR forecast.}\ The VaR-forecasts as well as the realized portfolio returns for all four confidence levels are shown in Figures \ref{fig:forecasts1} and \ref{fig:forecasts2}. In both figures, Panels (a) and (b) show the realized portfolio returns and the VaR forecasts computed by the use of the nonparametric vine copula and the parametric vine copula model calibrated via the heuristic based on Akaike's Information Criterion, respectively.

\begin{center}
--- insert Figures \ref{fig:forecasts1} and \ref{fig:forecasts2} here ---\\
\end{center}

The plots in Figures \ref{fig:forecasts1} and \ref{fig:forecasts2} show that both the nonparametric and the parametric model yield rather accurate forecasts of the portfolio's risk. While all VaR-forecasts are sufficiently close to the realized portfolio returns, exceedances of the VaR-forecasts occur only in case of large losses on the portfolio investment. Also, we can see that both the parametric and nonparametric model specifications yield quite similar VaR-forecasts. Thus, the proposed nonparametric vine copula model seems to perform exceptionally well even for low-dimensional portfolios. Furthermore, the finding of comparable VaR forecasts of both the nonparametric and parametric model remains valid for all four VaR confidence levels we consider. In addition to this, the results presented in Figures \ref{fig:forecasts1} and \ref{fig:forecasts2} also show that both models adequately adapt the VaR-forecasts to changes in the portfolio returns' volatility.\footnote{The flexible adjustment of both models to changes in return volatility also underlines the fact that a static dependence model in conjunction with dynamic marginal models suffices to model and forecast the dynamics of a multivariate return distribution.} To further assist in the interpretation of the results, Figures \ref{fig:exceed1} and \ref{fig:exceed2} highlight the positive and negative VaR-exceedances for all models and the three confidence levels for a long position in the portfolio. 

\begin{center}
--- insert Figures \ref{fig:exceed1} and \ref{fig:exceed2} here ---\\
\end{center}

Both figures underline our first impression from Figures \ref{fig:forecasts1} and \ref{fig:forecasts2} that both models forecast the VaR of the portfolio quite accurately. We can see from Figure \ref{fig:forecasts2} that not only do both models yield (approximately) correct numbers of VaR-exceedances for all three confidence levels, the exceedances also seem to occur randomly in time. Most importantly, however, our proposed nonparametric vine copula model with GARCH margins easily matches the heuristically calibrated parametric vine w.r.t. the accuracy of VaR-forecasting even for a relatively low-dimensional portfolio. To further substantiate this finding, we perform three formal backtests on the results of both the parametric and nonparametric vine models. The results of the three backtests are presented in Table \ref{tab:backtest}.

\begin{center}
--- insert Table \ref{tab:backtest} here ---\\
\end{center}

The backtesting results stress our finding that both models yield comparable results. For example, all but one VaR models cannot be rejected at the $99\%$ confidence level based on the test of conditional coverage. Although the results of the unconditional duration-based backtest imply a significantly worse forecasting accuracy of both models, the p-values for both the nonparametric and the parametric model are comparable for different confidence levels of the VaR. This indicates that neither model outperforms the other one based on our second backtest. If we use the conditional duration-based test of \citet{christoffersenpelletier:2004} instead, none of the VaR-models is rejected. Turning to the number of VaR-exceedances, the results of our nonparametric vine copula model are slightly better for the $5\%$-VaR than those of the parametric benchmark while the opposite is true for the (for most practical uses too optimistic and thus unsuitable) $10\%$-VaR.

Our backtesting results indicate that both models yield acceptable VaR-forecasts for a relatively low-dimensional portfolio. One could conclude from this finding that in general using our nonparametric vine copula model does not yield significantly better VaR-forecasts. However, one has to keep in mind that our empirical analysis was deliberately aimed at testing the hypothesis that the nonparametric model yields accurate VaR-forecasts even in lower dimensions. In unreported tests of high-dimensional portfolios, the parametric benchmark suffered from the same numerical instability that was also observed in our simulation study. At the same time, our nonparametric model produced accurate VaR-forecasts in a numerically stable fashion even for high-dimensional portfolios when the parametric benchmark had either broken down or produced woefully inaccurate VaR-forecasts.

\section{Summary}
In this paper, we propose to model the pair-copulas in a vine copula model nonparametrically by the use of Bernstein copulas. Our proposed model has the advantage of a significantly reduced model risk as it avoids the error-prone selection of pair-copulas from candidate parametric copula families. In contrast to previous studies on the use of discrete empirical copulas as pair-copulas, our proposed use of Bernstein copulas has the additional advantage that the building blocks in a vine model are approximated by smooth functions from which one can easily simulate random samples. We test the approximation error of the smooth nonparametric Bernstein vine copula model against a parametric benchmark calibrated by the use of a sequential heuristic based on AIC. The superiority of our proposed model is exemplified in an empirical risk management application.

The results we find in our simulation study show that for low-dimensional problems, the parametric modeling approach outperforms our proposed nonparametric approach only marginally. However, the differences in the approximation error quickly vanish for higher dimensions with both models yielding comparable average approximations errors for dimensions $d=13$ and higher. At the same time, our proposed nonparametric vine copula model does not suffer from numerical instability and error propagation which plagues the parametric benchmark due to an increasing number of wrongly selected parametric pair-copulas.

In the empirical risk management application, we test whether the differences in the average approximation error of the parametric and nonparametric vine copula models cause significant differences in both models' accuracy of forecasting the VaR of a low-dimensional asset portfolio. The results of our analysis show that even in lower dimensions ($d=5$), our nonparametric vine copula model yields VaR-forecasts that cannot be rejected by several different formal backtests. The proposed nonparametric vine copula model thus seems to match the (good) results of a parametric vine copula model in lower dimensions and significantly outperforms this benchmark in higher dimensions.

A natural extension of our model would be to consider more sophisticated smooth approximations of the empirical copula. Cubic B-splines and non-uniform rational B-splines (NURBS) appear as natural candidates for this job. While Bernstein copulas have been shown to be good smooth nonparametric replacements for parametric pair-copulas, spline copulas should yield even better approximations while at the same time yielding numerically stable vine model calibrations as well. We intend to analyze the suitability of spline copulas in vine models in future research.

\newpage
\singlespacing

\begin{thebibliography}{49}
\newcommand{\enquote}[1]{``#1''}
\expandafter\ifx\csname natexlab\endcsname\relax\def\natexlab#1{#1}\fi

\bibitem[\protect\citeauthoryear{Aas and Berg}{Aas and Berg}{2009}]{aas:2009b}
\textsc{Aas, K. and D.~Berg} (2009): \enquote{Models for construction of
  multivariate dependence - a comparison study,} \emph{European Journal of
  Finance}, 15, 639--659.

\bibitem[\protect\citeauthoryear{Aas, Czado, Frigessi, and Bakken}{Aas
  et~al.}{2009}]{aas:2009a}
\textsc{Aas, K., C.~Czado, A.~Frigessi, and H.~Bakken} (2009):
  \enquote{{Pair-Copula Constructions of Multiple Dependence},}
  \emph{Insurance: Mathematics and Economics}, 44, 182--198.

\bibitem[\protect\citeauthoryear{Ausin and Lopes}{Ausin and
  Lopes}{2010}]{ausin:2010}
\textsc{Ausin, M. and H.~Lopes} (2010): \enquote{Time-varying joint
  distribution through copulas,} \emph{Computational Statistics and Data
  Analysis}, 54(11), 2383--2399.

\bibitem[\protect\citeauthoryear{{Basel Committee on Banking
  Supervision}}{{Basel Committee on Banking Supervision}}{1996}]{Basel}
\textsc{{Basel Committee on Banking Supervision}} (1996): \enquote{Supervisory
  framework for the use of backtesting in conjunction with the internal models
  approach to market risk capital requirements,} Tech. rep., Bank of
  International Settlements.

\bibitem[\protect\citeauthoryear{Bedford and Cooke}{Bedford and
  Cooke}{2001}]{bedford:2001}
\textsc{Bedford, T. and R.~Cooke} (2001): \enquote{Probability density
  decomposition for conditionally dependent random variables modeled by vines,}
  \emph{Annals of Mathematics and Artificial Intelligence}, 32, 245--268.

\bibitem[\protect\citeauthoryear{Bedford and Cooke}{Bedford and
  Cooke}{2002}]{bedford:2002}
---\hspace{-.1pt}---\hspace{-.1pt}--- (2002): \enquote{{Vines - A New Graphical
  Model for Dependent Random Variables},} \emph{Annals of Statistics}, 30,
  1031--1068.

\bibitem[\protect\citeauthoryear{Berkowitz, Christoffersen, and
  Pelletier}{Berkowitz et~al.}{2011}]{Berk}
\textsc{Berkowitz, J., P.~Christoffersen, and D.~Pelletier} (2011):
  \enquote{Evaluating Value-at-Risk models with desk-level data,}
  \emph{Management Science}, 57, 2213--2227.

\bibitem[\protect\citeauthoryear{Brechmann and Czado}{Brechmann and
  Czado}{2012}]{brechczado:2012}
\textsc{Brechmann, E. and C.~Czado} (2012): \enquote{Risk management with
  high-dimensional vine copulas: An analysis of the Euro Stoxx 50,}
  \emph{working paper}.

\bibitem[\protect\citeauthoryear{Brechmann, Czado, and Aas}{Brechmann
  et~al.}{2012}]{brech:2012}
\textsc{Brechmann, E., C.~Czado, and K.~Aas} (2012): \enquote{Truncated Regular
  Vines in High Dimensions with Applications to Financial Data,} \emph{Canadian
  Journal of Statistics}, 40(1), 68--85.

\bibitem[\protect\citeauthoryear{Chan and Kroese}{Chan and Kroese}{2012}]{Chan}
\textsc{Chan, J.~C. and D.~P. Kroese} (2012): \enquote{{Efficient estimation of
  large portfolio loss probabilities in t-copula models},} \emph{European
  Journal of Operational Research}, 205, 361–367.

\bibitem[\protect\citeauthoryear{Christoffersen}{Christoffersen}{1998}]{chris1}
\textsc{Christoffersen, P.} (1998): \enquote{{Evaluating Interval Forecasts},}
  \emph{International Economic Reviewe}, 39, 841--862.

\bibitem[\protect\citeauthoryear{Christoffersen and Pelletier}{Christoffersen
  and Pelletier}{2004{\natexlab{a}}}]{chris2}
\textsc{Christoffersen, P. and D.~Pelletier} (2004{\natexlab{a}}):
  \enquote{{Backtesting Value-at-Risk: A Duration-Based Approach},}
  \emph{Journal of Financial Econometrics}, 2, 84--108.

\bibitem[\protect\citeauthoryear{Christoffersen and Pelletier}{Christoffersen
  and Pelletier}{2004{\natexlab{b}}}]{christoffersenpelletier:2004}
---\hspace{-.1pt}---\hspace{-.1pt}--- (2004{\natexlab{b}}):
  \enquote{{Backtesting Value-at-Risk: A duration-based approach},}
  \emph{Journal of Financial Econometrics}, 2, 84--108.

\bibitem[\protect\citeauthoryear{Deheuvels}{Deheuvels}{1979}]{deheuvels:1979}
\textsc{Deheuvels, P.} (1979): \enquote{{La fonction de d\'ependance empirique
  et ses propri\'et\'es - Un test non param\'etrique d'ind\'ependance},}
  \emph{Acad\'emie Royale de Belgique - Bulletin de la Classe des Sciences - 5e
  S\'erie}, 65, 274--292.

\bibitem[\protect\citeauthoryear{Deheuvels}{Deheuvels}{1981}]{deheuvels:1981}
---\hspace{-.1pt}---\hspace{-.1pt}--- (1981): \enquote{An asymptotic
  decomposition for multivariate distribution-free tests of independence,}
  \emph{J. Multivariate Anal.}, 11, 102--113.

\bibitem[\protect\citeauthoryear{Dias and Embrechts}{Dias and
  Embrechts}{2009}]{dias:2009}
\textsc{Dias, A. and P.~Embrechts} (2009): \enquote{Testing for structural
  changes in exchange rates' dependence beyond linear correlation,} \emph{The
  European Journal of Finance}, 15, 619--637.

\bibitem[\protect\citeauthoryear{Diers, Eling, and Marek}{Diers
  et~al.}{2012}]{diers:2012}
\textsc{Diers, D., M.~Eling, and S.~Marek} (2012): \enquote{Dependence modeling
  in non-life insurance using the Bernstein copula,} \emph{Insurance:
  Mathematics and Economics}, 50, 430--436.

\bibitem[\protect\citeauthoryear{Di\ss{}mann, Brechmann, Czado, and
  Kurowicka}{Di\ss{}mann et~al.}{forthcoming}]{dissmann:2011}
\textsc{Di\ss{}mann, J., E.~Brechmann, C.~Czado, and D.~Kurowicka}
  (forthcoming): \enquote{Selecting and estimating regular vine copulae and
  application to financial returns,} \emph{Computational Statistics and Data
  Analysis}.

\bibitem[\protect\citeauthoryear{Embrechts, McNeil, and Straumann}{Embrechts
  et~al.}{2002}]{embrechts:2002}
\textsc{Embrechts, P., A.~McNeil, and D.~Straumann} (2002):
  \enquote{Correlation and dependence in risk management: properties and
  pitfalls,} in \emph{Risk Management: Value at Risk and Beyond}, ed. by
  M.~Dempster, Cambridge University Press, Cambridge, 176--223.

\bibitem[\protect\citeauthoryear{Engle and Russell}{Engle and
  Russell}{1998}]{engle:1998}
\textsc{Engle, R. and J.~Russell} (1998): \enquote{Autoregressive Conditional
  Duration: A New Model for Irregularly Spaced Transaction Data,}
  \emph{Econometrica}, 66, 1127--1162.

\bibitem[\protect\citeauthoryear{Escanciano and Pei}{Escanciano and
  Pei}{2012}]{escanciano}
\textsc{Escanciano, J.~C. and P.~Pei} (2012): \enquote{{Pitfalls in backtesting
  Historical Simulation VaR models},} \emph{Journal of Banking and Finance},
  36, 2233--2244.

\bibitem[\protect\citeauthoryear{Fantazzini}{Fantazzini}{2009}]{fantazzini:2009b}
\textsc{Fantazzini, D.} (2009): \enquote{{Market Risk Management for Emerging
  Markets: Evidence from the Russian Stock Market},} in \emph{{Emerging
  markets: Performance, analysis and innovation}}, ed. by G.~Gregoriou, Chapman
  and Hall / CRC Finance, 533--554.

\bibitem[\protect\citeauthoryear{Fischer, K\"ock, Schl\"uter, and
  Weigert}{Fischer et~al.}{2009}]{fischer:2009}
\textsc{Fischer, M., C.~K\"ock, S.~Schl\"uter, and F.~Weigert} (2009):
  \enquote{An empirical analysis of multivariate copula model,}
  \emph{Quantitative Finance}, 7, 839--854.

\bibitem[\protect\citeauthoryear{Genest, Gendron, and Bourdeau-Brien}{Genest
  et~al.}{2009{\natexlab{a}}}]{genest:2009}
\textsc{Genest, C., M.~Gendron, and M.~Bourdeau-Brien} (2009{\natexlab{a}}):
  \enquote{The advent of copulas in finance,} \emph{European Journal of
  Finance}, 15, 609--618.

\bibitem[\protect\citeauthoryear{Genest, R\'emillard, and Beaudoin}{Genest
  et~al.}{2009{\natexlab{b}}}]{genestelal:2009}
\textsc{Genest, C., B.~R\'emillard, and D.~Beaudoin} (2009{\natexlab{b}}):
  \enquote{Goodness-of-fit tests for copulas: A review and a power study,}
  \emph{Insurance: Mathematics and Economics}, 44, 199--213.

\bibitem[\protect\citeauthoryear{Goldfarb and Idnani}{Goldfarb and
  Idnani}{1982}]{goldfarb:1982}
\textsc{Goldfarb, D. and A.~Idnani} (1982): \enquote{Dual and Primal-Dual
  Methods for Solving Strictly Convex Quadratic Programs,} Springer Verlag,
  Berlin, 226--239.

\bibitem[\protect\citeauthoryear{Goldfarb and Idnani}{Goldfarb and
  Idnani}{1983}]{goldfarb:1983}
---\hspace{-.1pt}---\hspace{-.1pt}--- (1983): \enquote{A numerically stable
  dual method for solving strictly convex quadratic programs,}
  \emph{Mathematical Programming}, 27, 1--33.

\bibitem[\protect\citeauthoryear{Gr{\o}nneberg and Hjort}{Gr{\o}nneberg and
  Hjort}{2012}]{gronneberg:2012}
\textsc{Gr{\o}nneberg, S. and N.~L. Hjort} (2012): \enquote{The Copula
  Information Criterion,} \emph{Scandinavian Journal of Statistics},
  forthcoming.

\bibitem[\protect\citeauthoryear{Grundke and Polle}{Grundke and
  Polle}{2012}]{grundke:2012}
\textsc{Grundke, P. and S.~Polle} (2012): \enquote{Crisis and risk
  dependencies,} \emph{European Journal of Operational Research}, 223,
  518--528.

\bibitem[\protect\citeauthoryear{Hafner and Reznikova}{Hafner and
  Reznikova}{2010}]{hafner:2010}
\textsc{Hafner, C. and O.~Reznikova} (2010): \enquote{Efficient estimation of a
  semiparametric dynamic copula models,} \emph{Computational Statistics and
  Data Analysis}, 54(11), 2069--2627.

\bibitem[\protect\citeauthoryear{Hansen and Lunde}{Hansen and
  Lunde}{2005}]{hansen:2005}
\textsc{Hansen, P. and A.~Lunde} (2005): \enquote{{A Forecast Comparison of
  Volatility Models: Does Anything Beat a GARCH(1,1)?}} \emph{Journal of
  Applied Econometrics}, 20, 873--889.

\bibitem[\protect\citeauthoryear{Hob{\ae}k-Haff and Segers}{Hob{\ae}k-Haff and
  Segers}{2012}]{hobaekhaff:2012}
\textsc{Hob{\ae}k-Haff, I. and J.~Segers} (2012): \enquote{Non-parametric
  estimation of pair-copula constructions with the empirical pair-copula,}
  \emph{Working Paper, January 2012}.

\bibitem[\protect\citeauthoryear{Hsu, Huang, and Chiou}{Hsu
  et~al.}{2011}]{hsu:2011}
\textsc{Hsu, C., C.~Huang, and W.~Chiou} (2011): \enquote{Effectiveness of
  Copula-Extreme Value Theory in Estimating Value-at-Risk: Empirical Evidence
  from Asian Emerging Markets,} \emph{Review of Quantitative Finance and
  Accounting}, Forthcoming.

\bibitem[\protect\citeauthoryear{Ince and Porter}{Ince and
  Porter}{2006}]{inceporter:2006}
\textsc{Ince, O. and R.~Porter} (2006): \enquote{{Individual Equity Return Data
  From Thomson Datastream: Handle With Care!}} \emph{Journal of Financial
  Research}, 29, 463--479.

\bibitem[\protect\citeauthoryear{Joe}{Joe}{1996}]{joe:1996}
\textsc{Joe, H.} (1996): \enquote{Families of m-variate distributions with
  given margins and m(m-1)/2 bivariate dependence parameters,} in
  \emph{Distributions with Fixed Marginals and Related Topics}, ed. by
  L.~R�schendorf, B.~Schweizer, and M.~Taylor, Hayward, CA: IMS Lecture Notes
  - Monograph Series, 120--141.

\bibitem[\protect\citeauthoryear{Joe}{Joe}{1997}]{joe:1997}
---\hspace{-.1pt}---\hspace{-.1pt}--- (1997): \emph{{Multivariate Models and
  Dependence Concepts}}, Chapman \&\ Hall, London.

\bibitem[\protect\citeauthoryear{Jondeau and Rockinger}{Jondeau and
  Rockinger}{2006}]{jondeau:2006}
\textsc{Jondeau, E. and M.~Rockinger} (2006): \enquote{{The copula-GARCH model
  of conditional dependencies: An international stock market application},}
  \emph{Journal of International Money and Finance}, 25(5), 827--853.

\bibitem[\protect\citeauthoryear{Kim, Silvapulle, and Silvapulle}{Kim
  et~al.}{2007}]{kim:2007}
\textsc{Kim, G., M.~Silvapulle, and P.~Silvapulle} (2007): \enquote{Comparison
  of semiparametric and parametric methods for estimating copulas,}
  \emph{Computational Statistics and Data Analysis}, 51, 2836--2850.

\bibitem[\protect\citeauthoryear{Kole, Koedijk, and Verbeek}{Kole
  et~al.}{2007}]{kole:2007}
\textsc{Kole, E., K.~Koedijk, and M.~Verbeek} (2007): \enquote{{Selecting
  Copulas for Risk Management},} \emph{Journal of Banking \& Finance}, 31,
  2405--2423.

\bibitem[\protect\citeauthoryear{Kulpa}{Kulpa}{1999}]{kulpa:1999}
\textsc{Kulpa, T.} (1999): \enquote{On approximation of copulas,}
  \emph{International Journal of Mathematics and Mathematical Sciences}, 22,
  259--269.

\bibitem[\protect\citeauthoryear{Kurowicka}{Kurowicka}{2010}]{kurowicka:2010}
\textsc{Kurowicka, D.} (2010): \enquote{Optimal truncation of vines,} in
  \emph{Dependence Modeling: Handbook on Vine Copulae}, ed. by D.~Kurowicka and
  H.~Joe, World Scientific Publishing Co.

\bibitem[\protect\citeauthoryear{Nikoloulopoulos, Joe, and Li}{Nikoloulopoulos
  et~al.}{2011}]{nikoloulopoulos:2011}
\textsc{Nikoloulopoulos, A., H.~Joe, and H.~Li} (2011): \enquote{Vine copulas
  with asymmetric tail dependence and applications to financial return data,}
  \emph{Computational Statistics and Data Analysis}, forthcoming.

\bibitem[\protect\citeauthoryear{Pfeifer, Strassburger, and Philipps}{Pfeifer
  et~al.}{2009}]{pfeifer:2009}
\textsc{Pfeifer, D., D.~Strassburger, and J.~Philipps} (2009):
  \enquote{Modelling and simulation of dependence structures in nonlife
  insurance with Bernstein copulas,} \emph{Working Paper, Carl von Ossietzky
  University, Oldenburg}.

\bibitem[\protect\citeauthoryear{Sancetta and Satchell}{Sancetta and
  Satchell}{2004}]{sancetta:2004}
\textsc{Sancetta, A. and S.~Satchell} (2004): \enquote{The Bernstein copula and
  its applications to modeling and approximations of multivariate
  distributions,} \emph{Econometric Theory}, 20, 1--38.

\bibitem[\protect\citeauthoryear{Shen, Y-Zhu, and Song}{Shen
  et~al.}{2008}]{shen:2008}
\textsc{Shen, X., Y-Zhu, and L.~Song} (2008): \enquote{{Linear B-spline copulas
  with applications to nonparametric estimation of copulas},}
  \emph{Computational Statistics and Data Analysis}, 52, 3806--3819.

\bibitem[\protect\citeauthoryear{Wei\ss}{Wei\ss}{2011}]{weiss:2011}
\textsc{Wei\ss, G.} (2011): \enquote{{Are Copula-GoF-tests of any practical
  use? Empirical evidence for stocks, commodities and FX futures},} \emph{The
  Quarterly Review of Economics and Finance}, 51(2), 173--188.

\bibitem[\protect\citeauthoryear{Wei\ss}{Wei\ss}{2012}]{weiss:2012rqfa}
---\hspace{-.1pt}---\hspace{-.1pt}--- (2012): \enquote{{Copula-GARCH vs.
  Dynamic Conditional Correlation - an empirical study on VaR and ES
  forecasting accuracy},} \emph{Review of Quantitative Finance and Accounting},
  forthcoming.

\bibitem[\protect\citeauthoryear{Whelan}{Whelan}{2004}]{whelan:2004}
\textsc{Whelan, N.} (2004): \enquote{{Sampling from Archimedean copulas},}
  \emph{Quantitative Finance}.

\bibitem[\protect\citeauthoryear{Yea, Liub, and Miaoa}{Yea et~al.}{2012}]{yea}
\textsc{Yea, W., C.~Liub, and B.~Miaoa} (2012): \enquote{{Measuring the
  subprime crisis contagion: Evidence of change point analysis of copula
  functions},} \emph{European Journal of Operational Research}, 222, 96--103.

\end{thebibliography}
\bibliographystyle{ecta}

\newpage
\section*{Figures and Tables}

\newpage

\newpage

\newpage

\newpage

\newpage
\begin{sidewaysfigure}[htbp]

\begin{center}
\includegraphics[width=11.2cm]{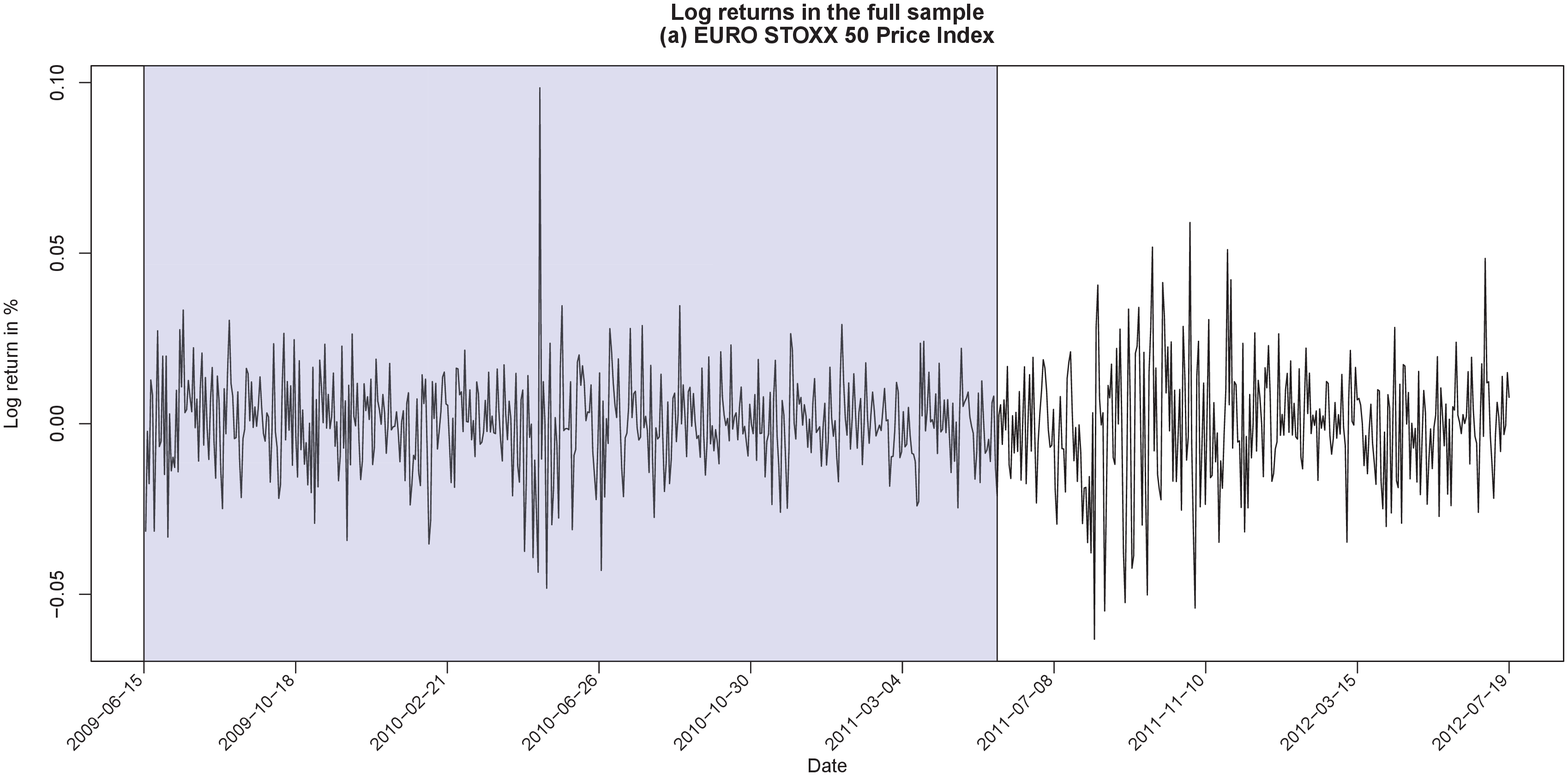}
\includegraphics[width=11.2cm]{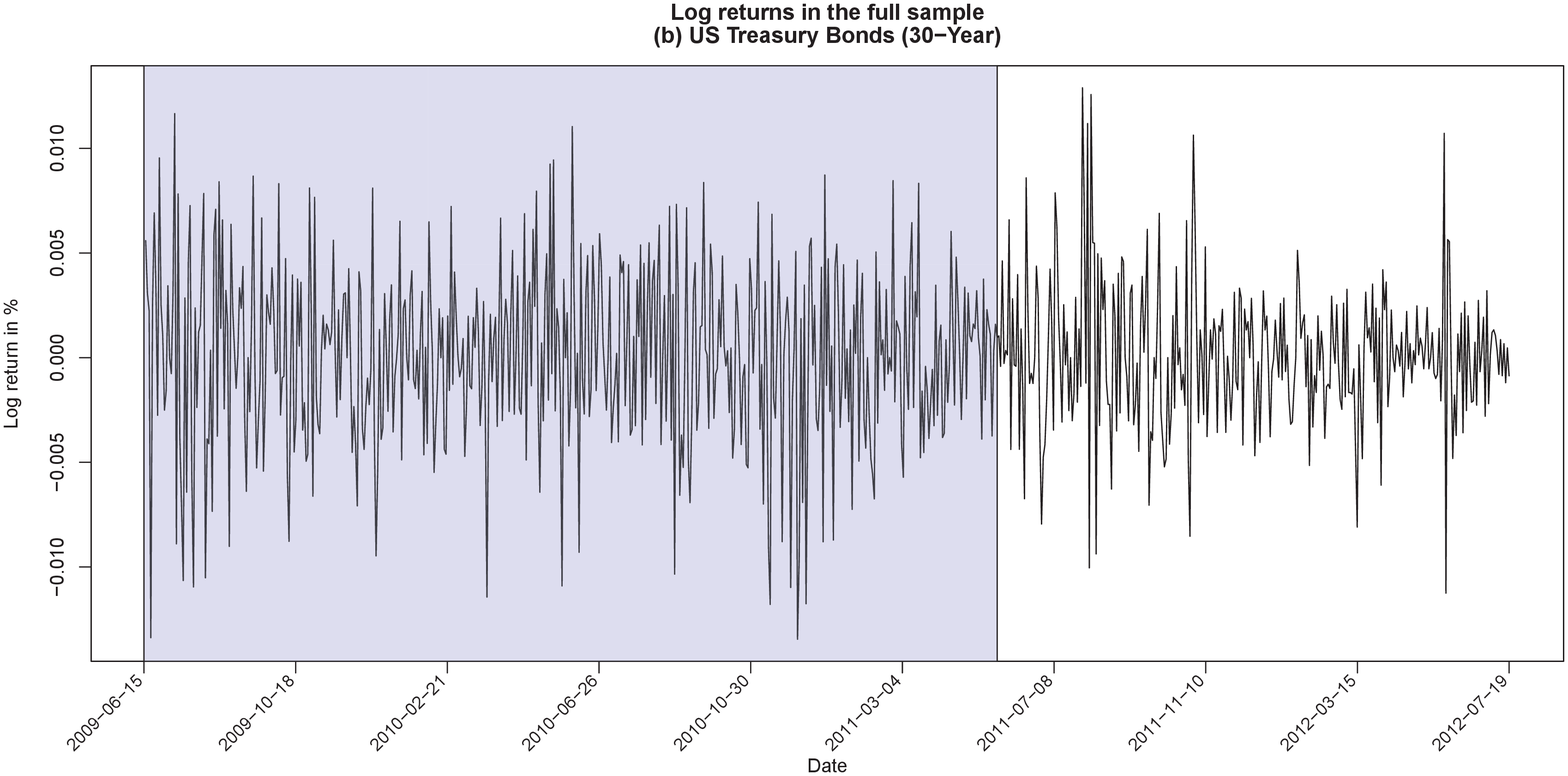}
\includegraphics[width=11.2cm]{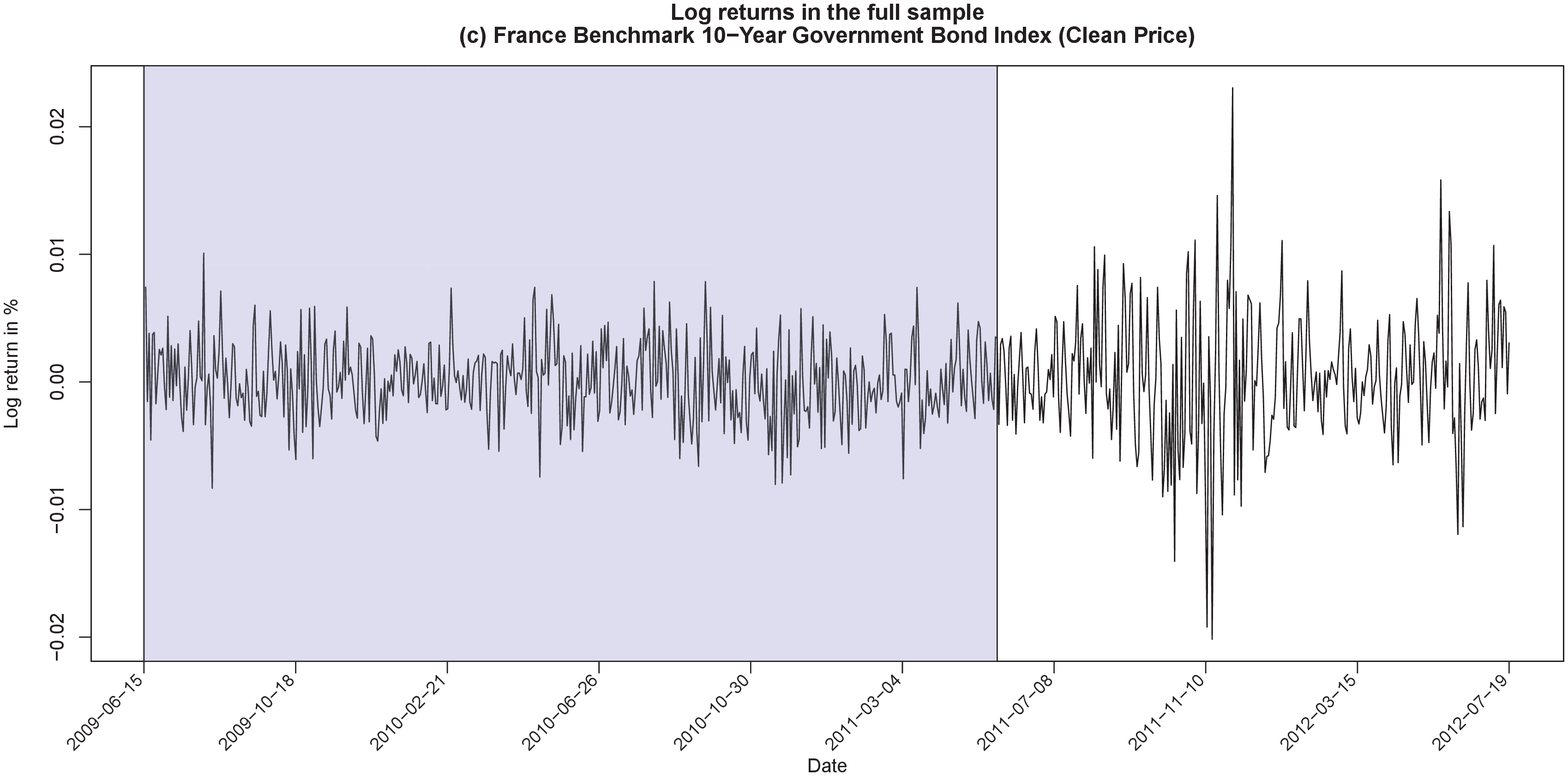}
\includegraphics[width=11.2cm]{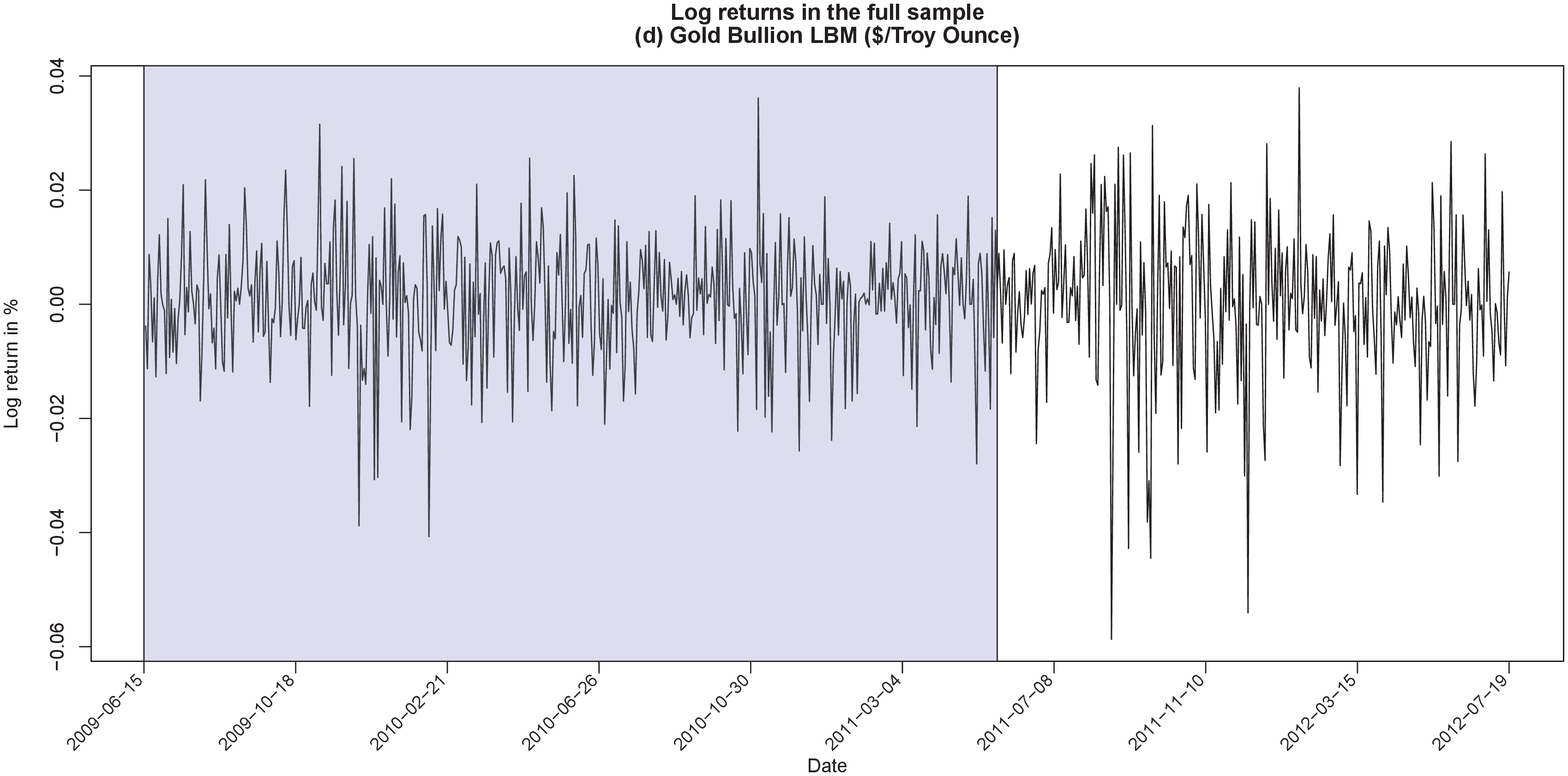}
\end{center}
\caption{{\bf Time series plots of log returns in the full sample.} The figure shows plots of the log returns on the EURO STOXX 50 Price Index in Panel (a), US Treasury Bonds (30-year) in Panel (b), France Benchmark 10-Year Government Bond Index (Clean Price) in Panel (c), Gold Bullion LBM (\$/Troy Ounce) in Panel (d), Crude Oil-Brent one-month forward (\$/BBL) in Panel (e) and the returns on an equal-weighted portfolio consisting of the five individual assets in Panel (f). The sample covers the period from June 15, 2009 to July 19, 2012 (800 trading days). The plots show the log returns during our complete sample and are divided into the initial in-sample of 500 trading days (shaded in grey) and the out-of-sample of 300 trading days.}
\label{fig:Hist1}
\end{sidewaysfigure}

\setcounter{figure}{2}

\begin{sidewaysfigure}[htbp]
\label{fig:Hist1continued}
\begin{center}
\includegraphics[width=15.0cm]{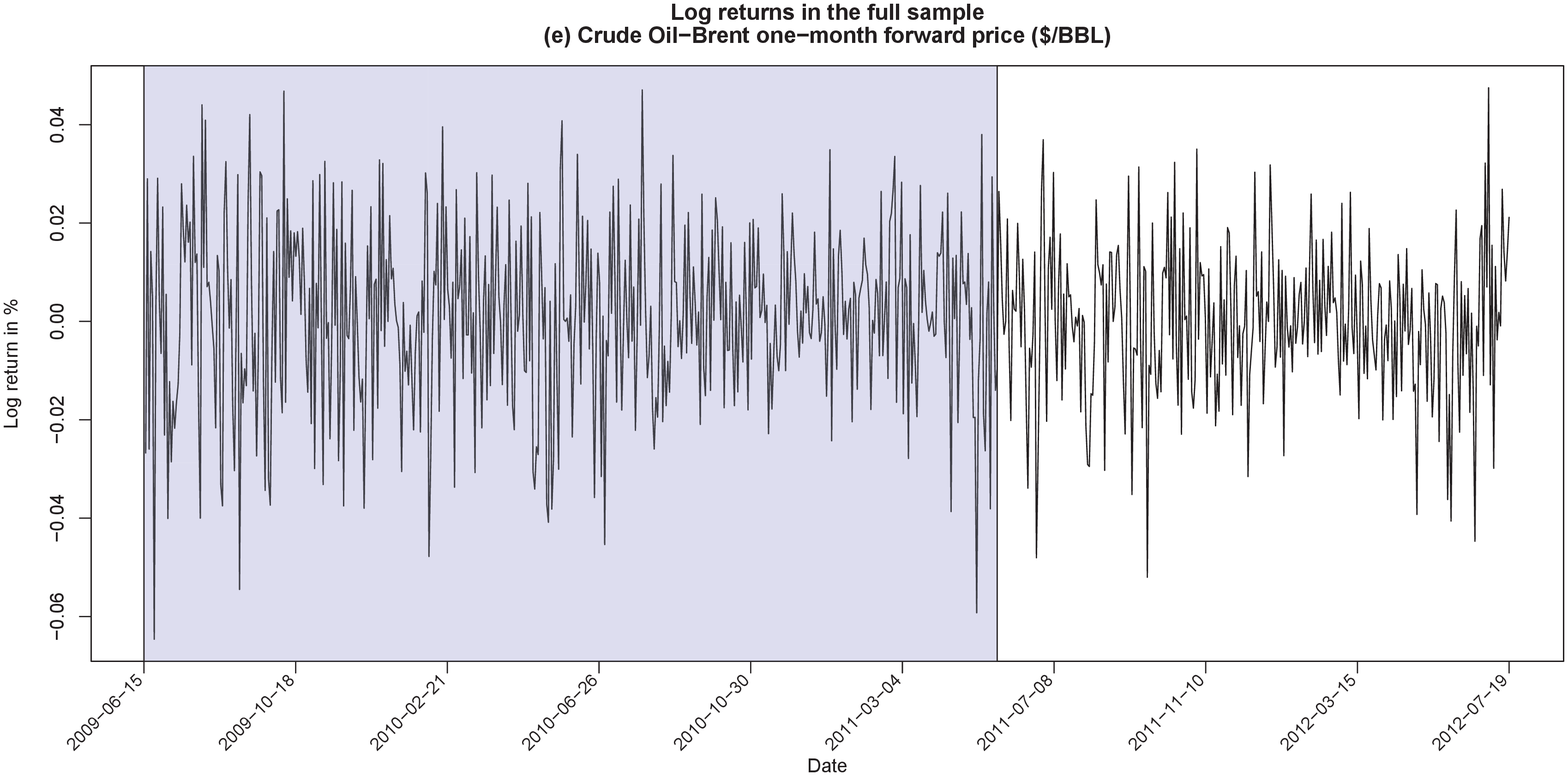}
\includegraphics[width=15.0cm]{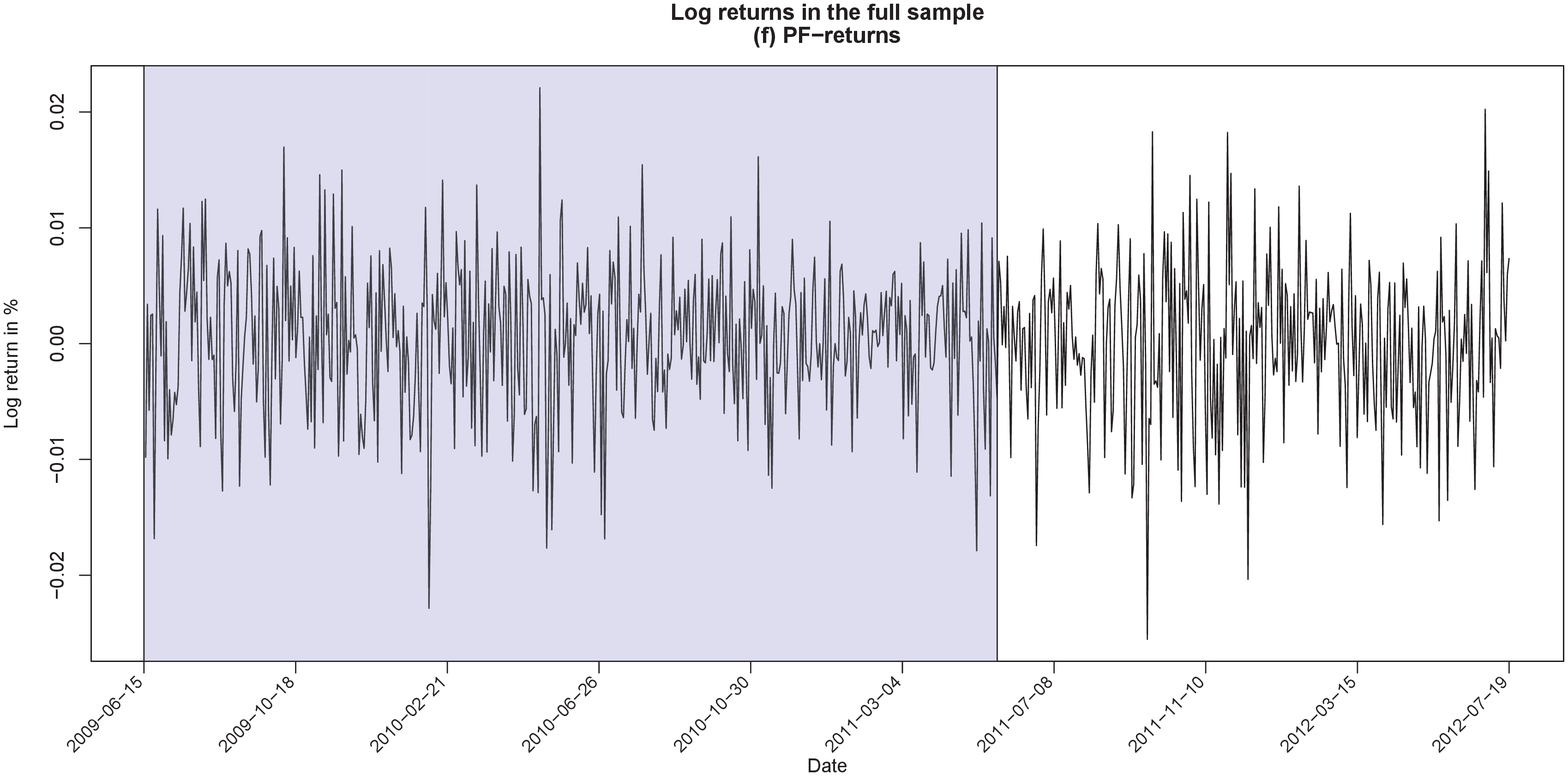}
\end{center}
\caption{{\bf Continued.}}
\end{sidewaysfigure}

\newpage
\begin{sidewaysfigure}[htbp]

\begin{center}
\includegraphics[width=17.0cm]{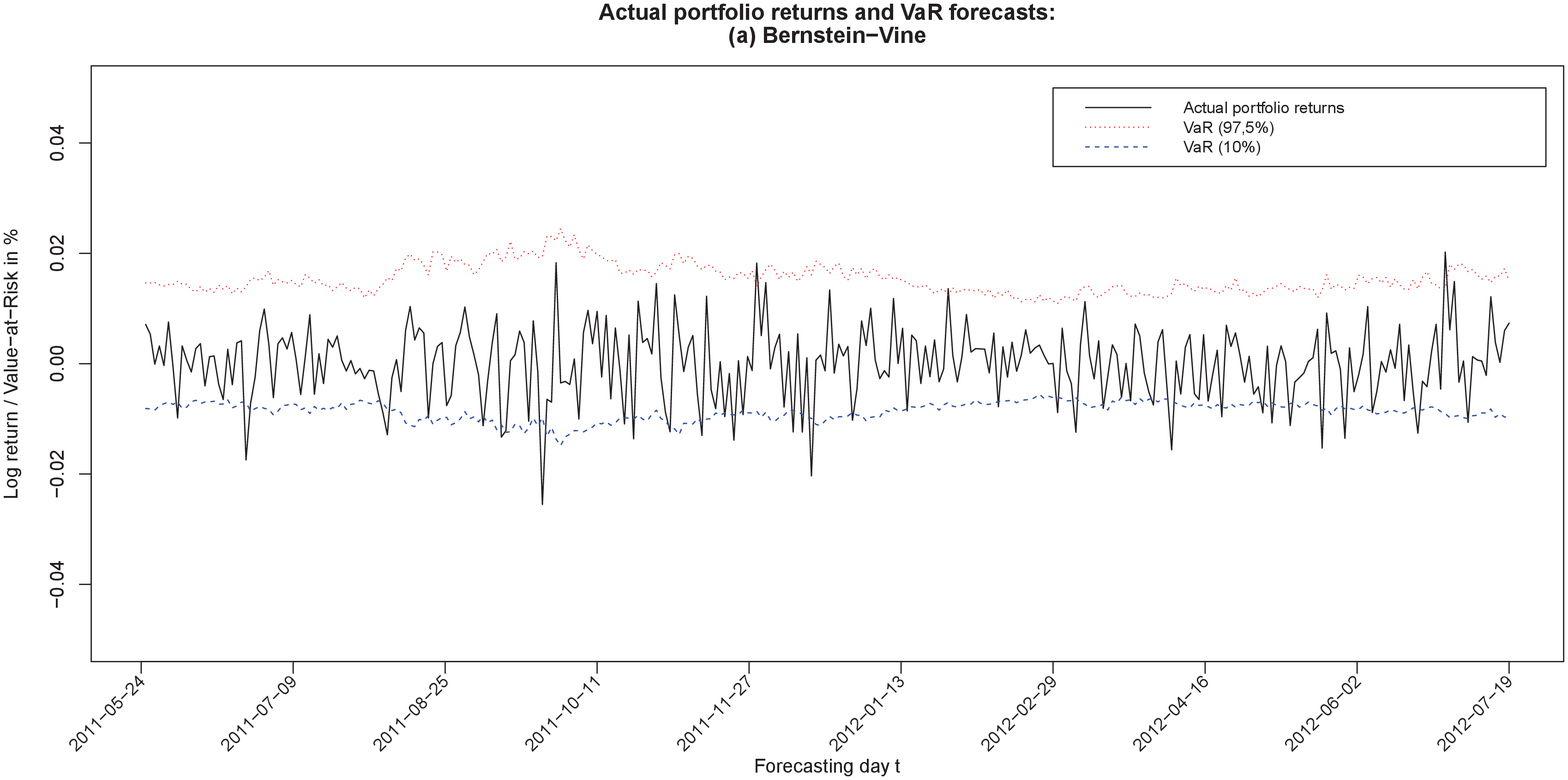}
\includegraphics[width=17.0cm]{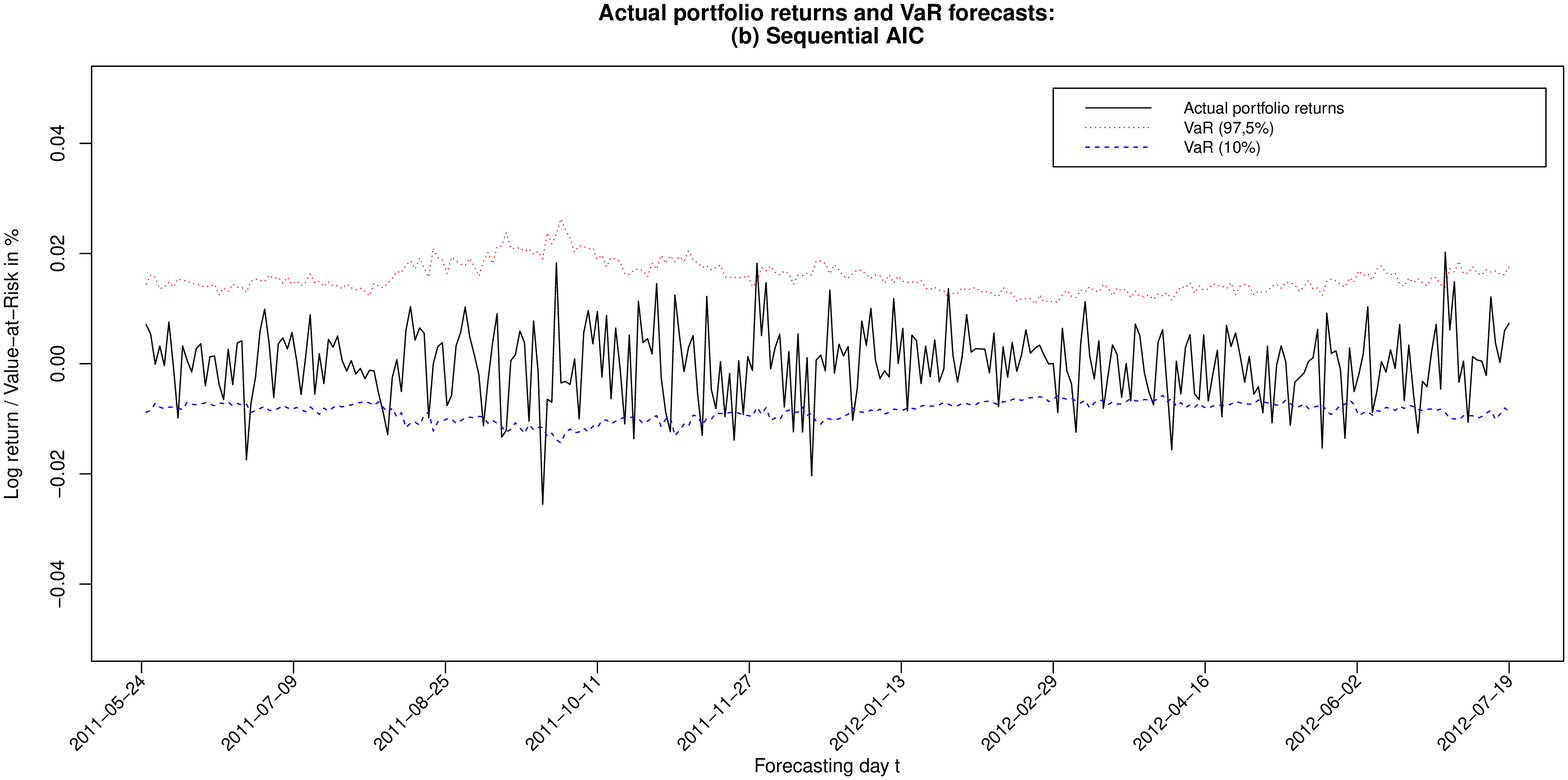}
\end{center}
\caption{{\bf Comparison of $10\%-$ and $97.5\%-$VaR forecasts and realized portfolio returns.} The figure shows plots of the log returns on the five-dimensional portfolio we consider in our empirical study and the VaR forecasts. Panel (a) presents the realized portfolio returns and the VaR forecasts computed by the use of the nonparametric vine copula as the dependence model. Panel (b) shows a corresponding comparison of the portfolio returns and the VaR forecasts estimated via a vine copula with the parametric pair-copulas chosen according to the sequential heuristic taken from the R-package \textit{CDVine} based on Akaike's Information Criterion. For both models, the $(1-\alpha)$-VaR is computed for confidence levels $\alpha\in\left\{10\%;97.5\%\right\}$. Both plots show results for the out-of-sample of $300$ trading days. The portfolio consists of the returns on the EURO STOXX 50 Price Index, US Treasury Bonds (30-year), France Benchmark 10-Year Government Bond Index, Gold Bullion LBM and Crude Oil-Brent one-month forward.}
\label{fig:forecasts1}
\end{sidewaysfigure}

\newpage
\begin{sidewaysfigure}[htbp]

\begin{center}
\includegraphics[width=17.0cm]{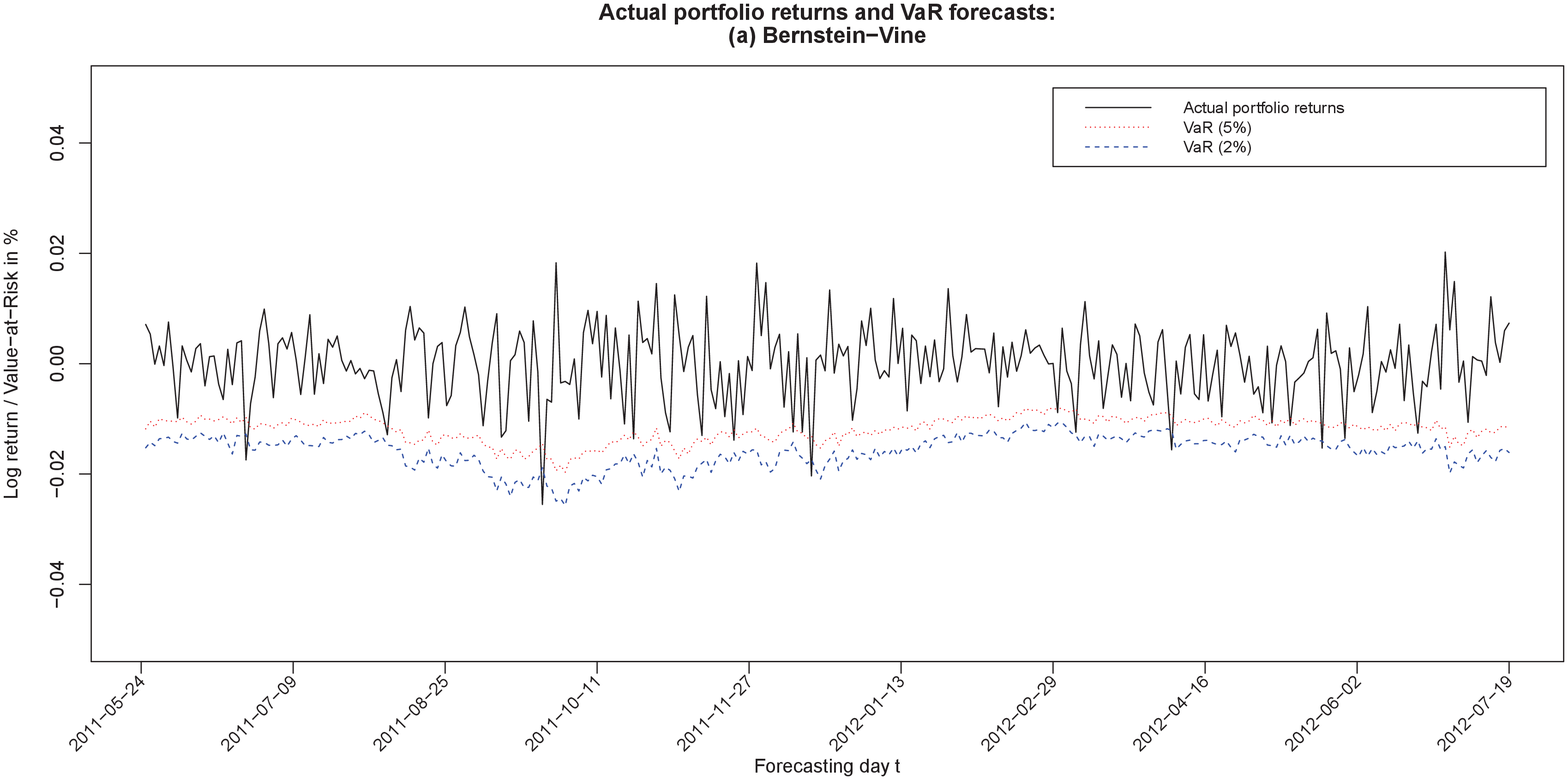}
\includegraphics[width=17.0cm]{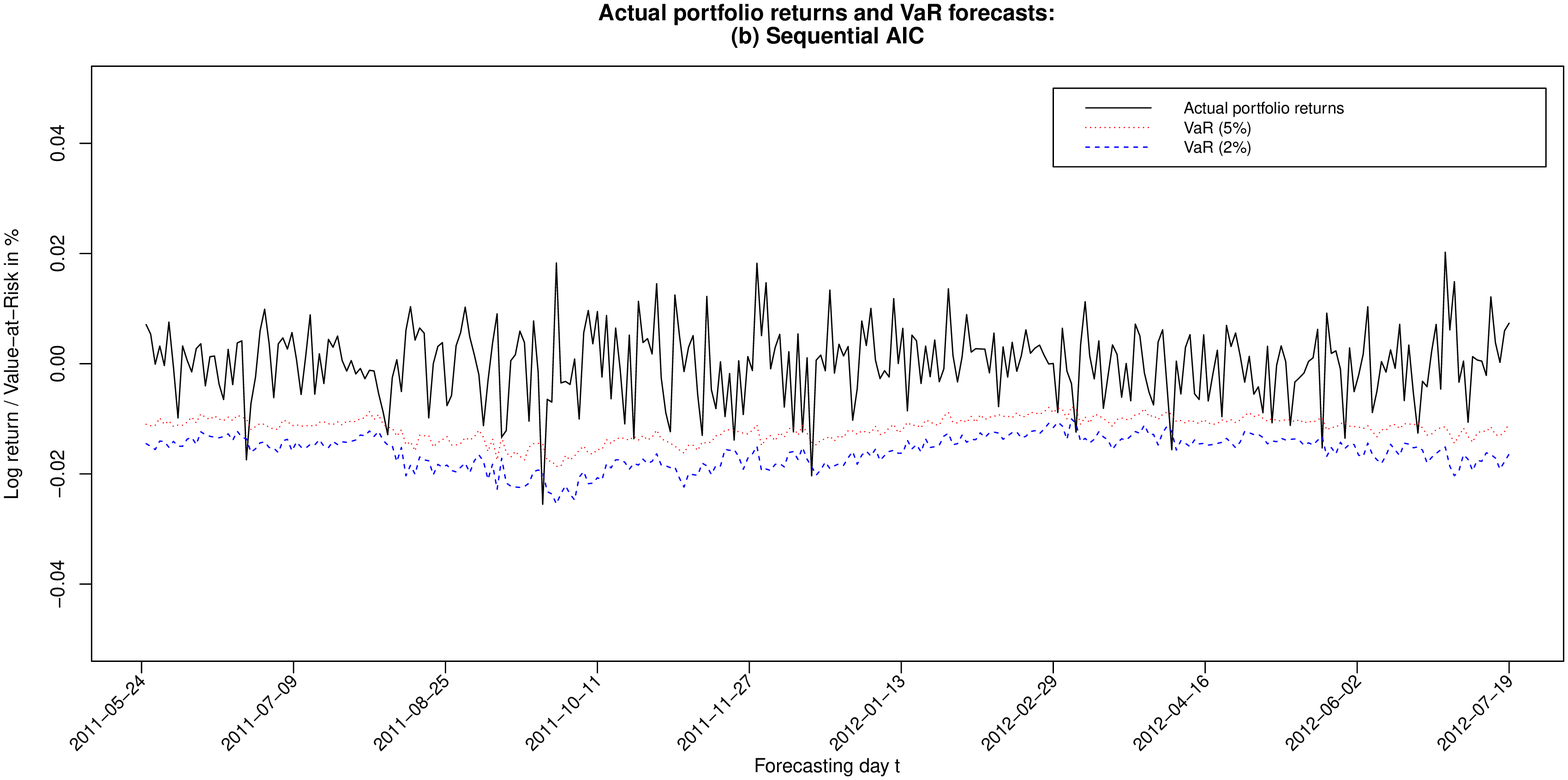}
\end{center}
\caption{{\bf Comparison of $2\%-$ and $5\%-$VaR forecasts and realized portfolio returns.} The figure shows plots of the log returns on the five-dimensional portfolio we consider in our empirical study and the VaR forecasts. Panel (a) presents the realized portfolio returns and the VaR forecasts computed by the use of the nonparametric vine copula as the dependence model. Panel (b) shows a corresponding comparison of the portfolio returns and the VaR forecasts estimated via a vine copula with the parametric pair-copulas chosen according to the sequential heuristic taken from the R-package \textit{CDVine} based on Akaike's Information Criterion. For both models, the $(1-\alpha)$-VaR is computed for confidence levels $\alpha\in\left\{2\%;5\%\right\}$. Both plots show results for the out-of-sample of $300$ trading days. The portfolio consists of the returns on the EURO STOXX 50 Price Index, US Treasury Bonds (30-year), France Benchmark 10-Year Government Bond Index, Gold Bullion LBM and Crude Oil-Brent one-month forward.}
\label{fig:forecasts2}
\end{sidewaysfigure}

\newpage
\begin{sidewaysfigure}[htbp]

\begin{center}
\includegraphics[width=11.2cm]{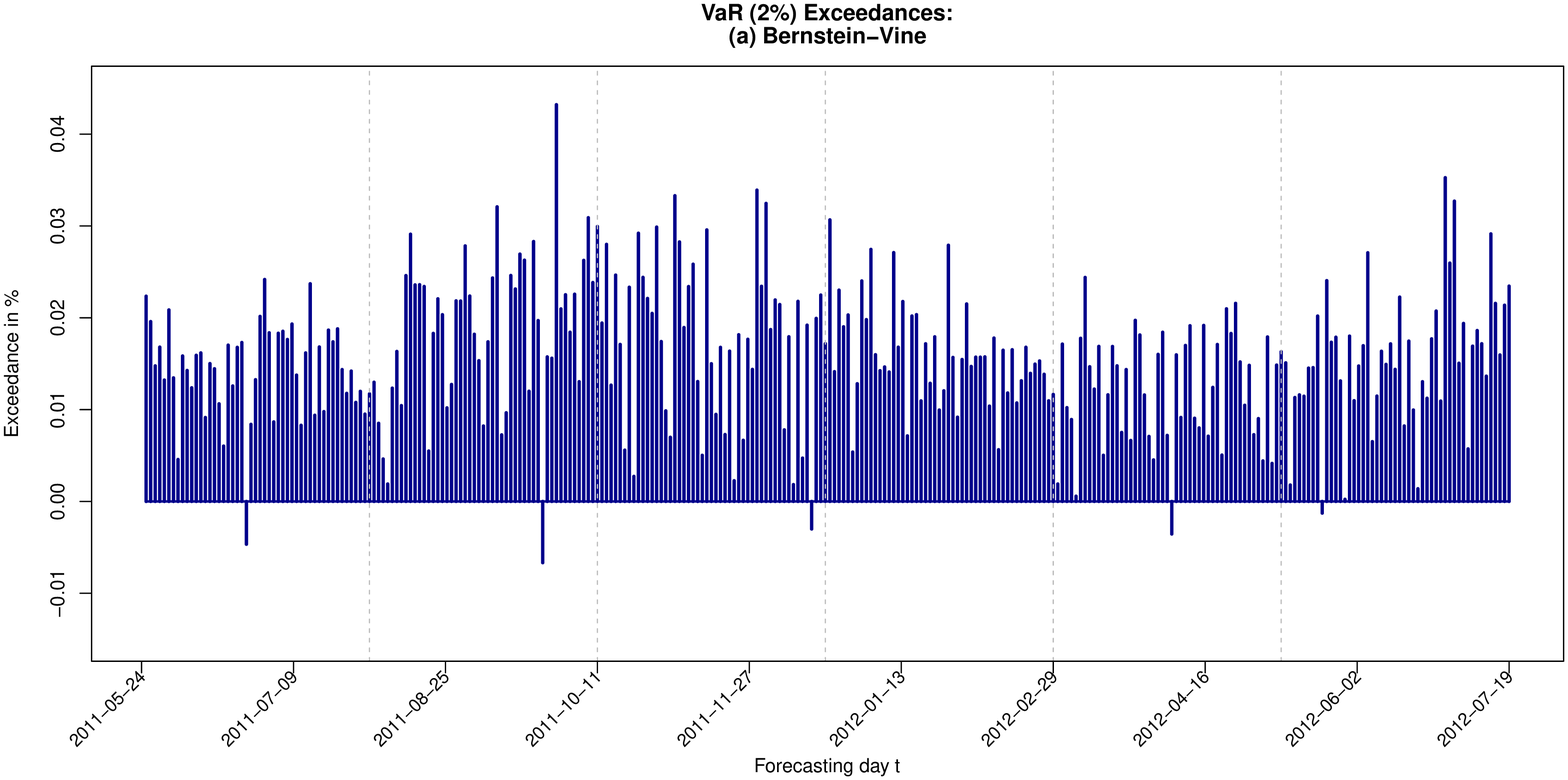}
\includegraphics[width=11.2cm]{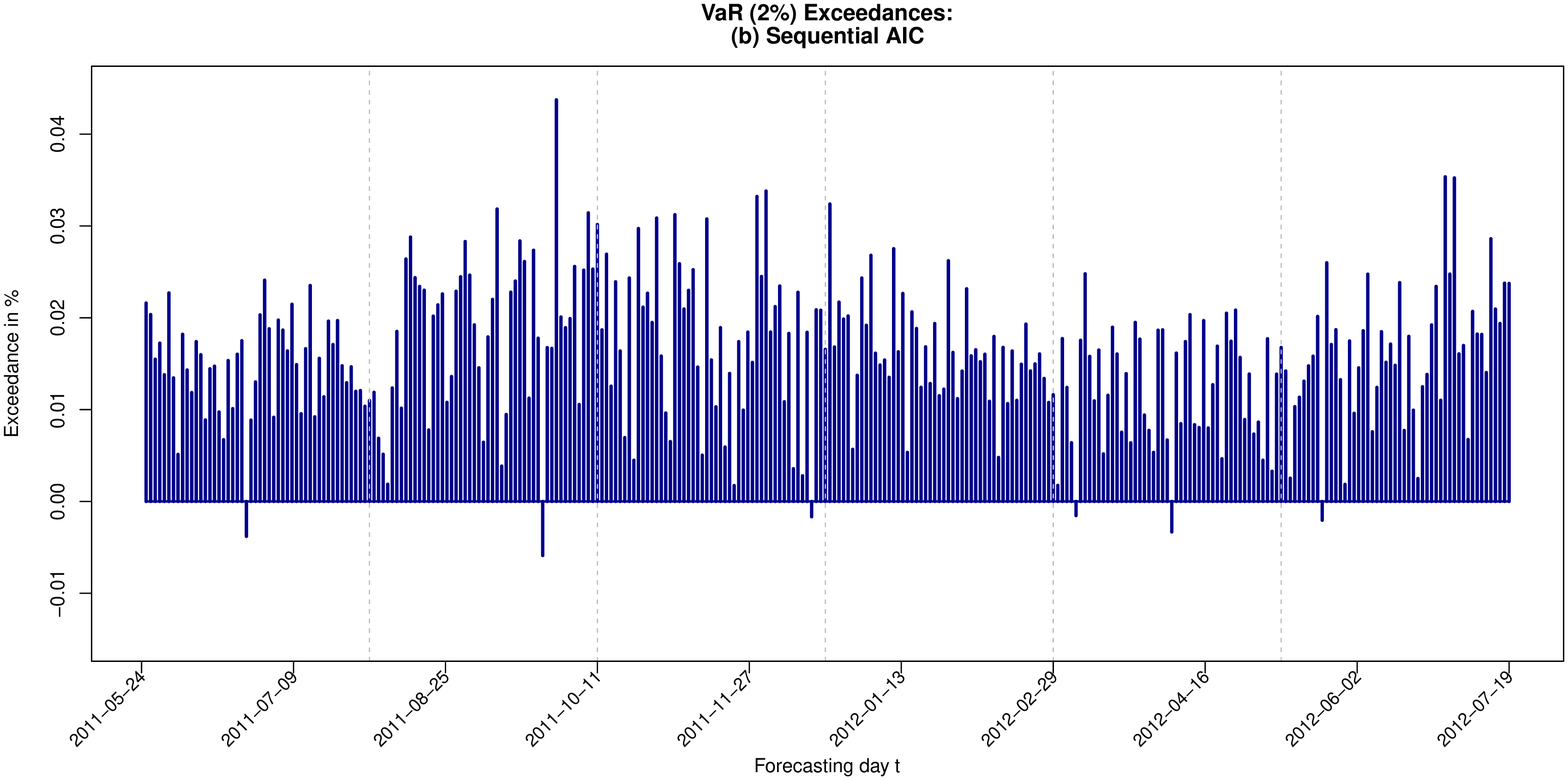}
\includegraphics[width=11.2cm]{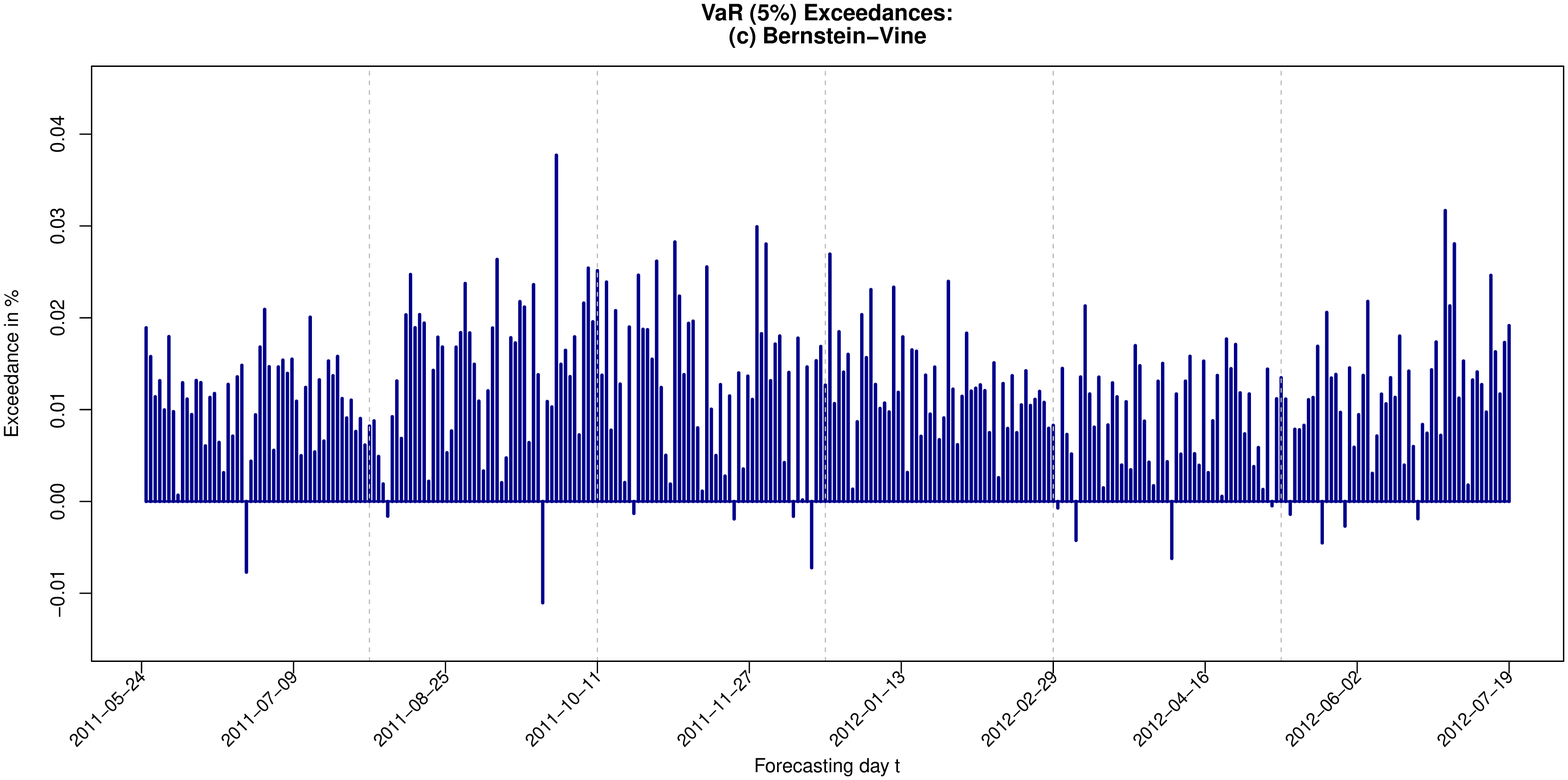}
\includegraphics[width=11.2cm]{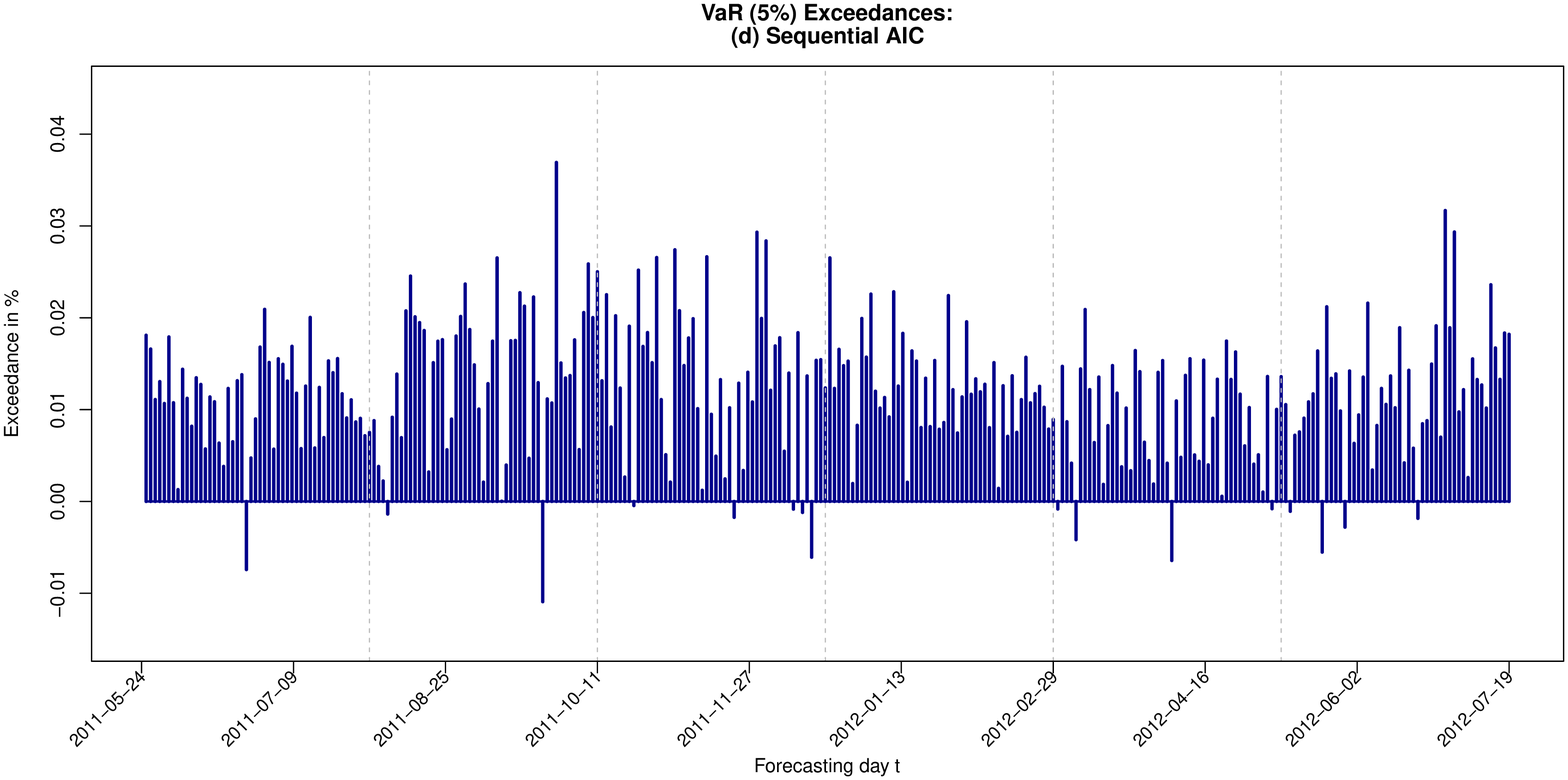}
\includegraphics[width=11.2cm]{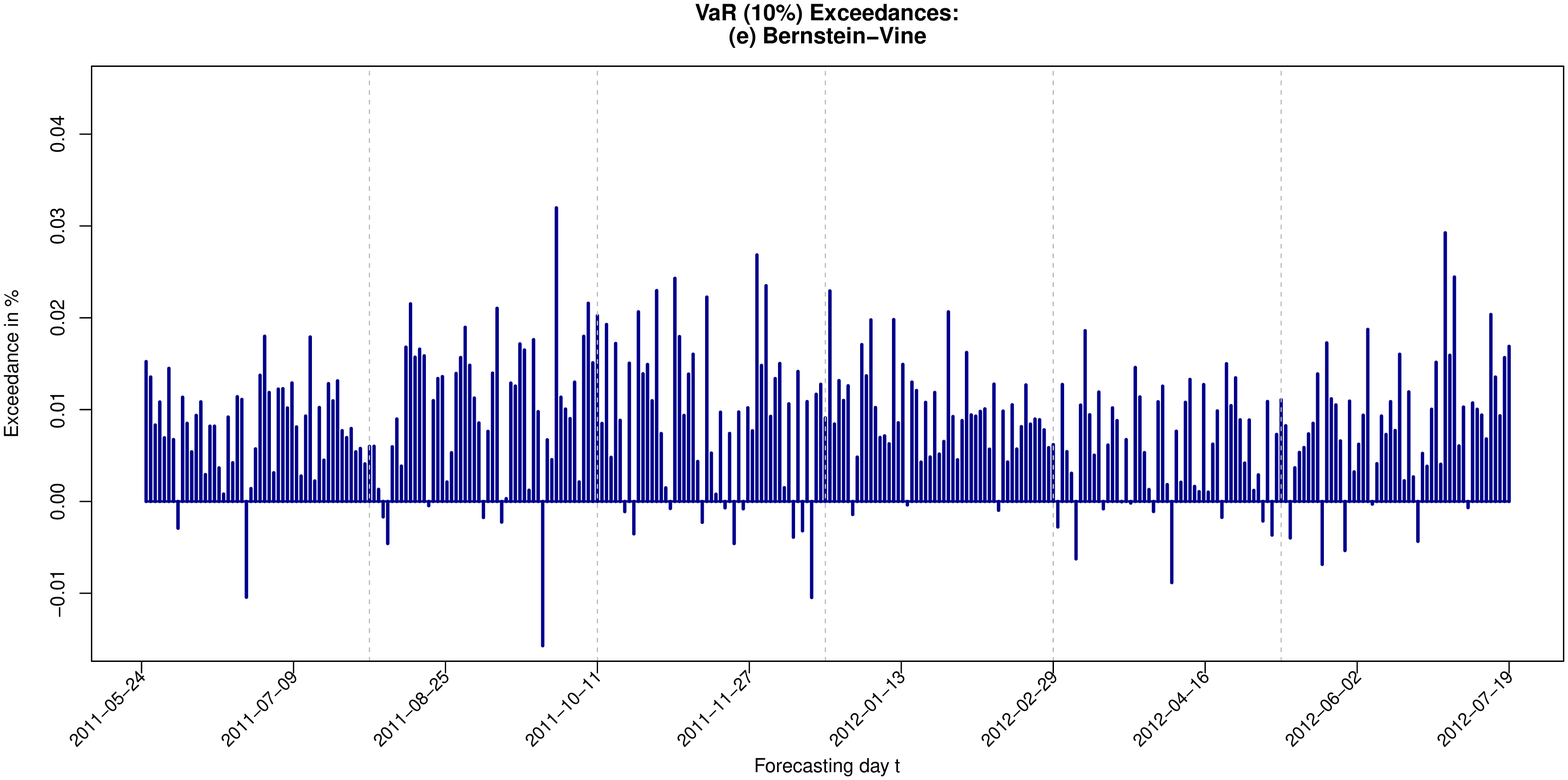}
\includegraphics[width=11.2cm]{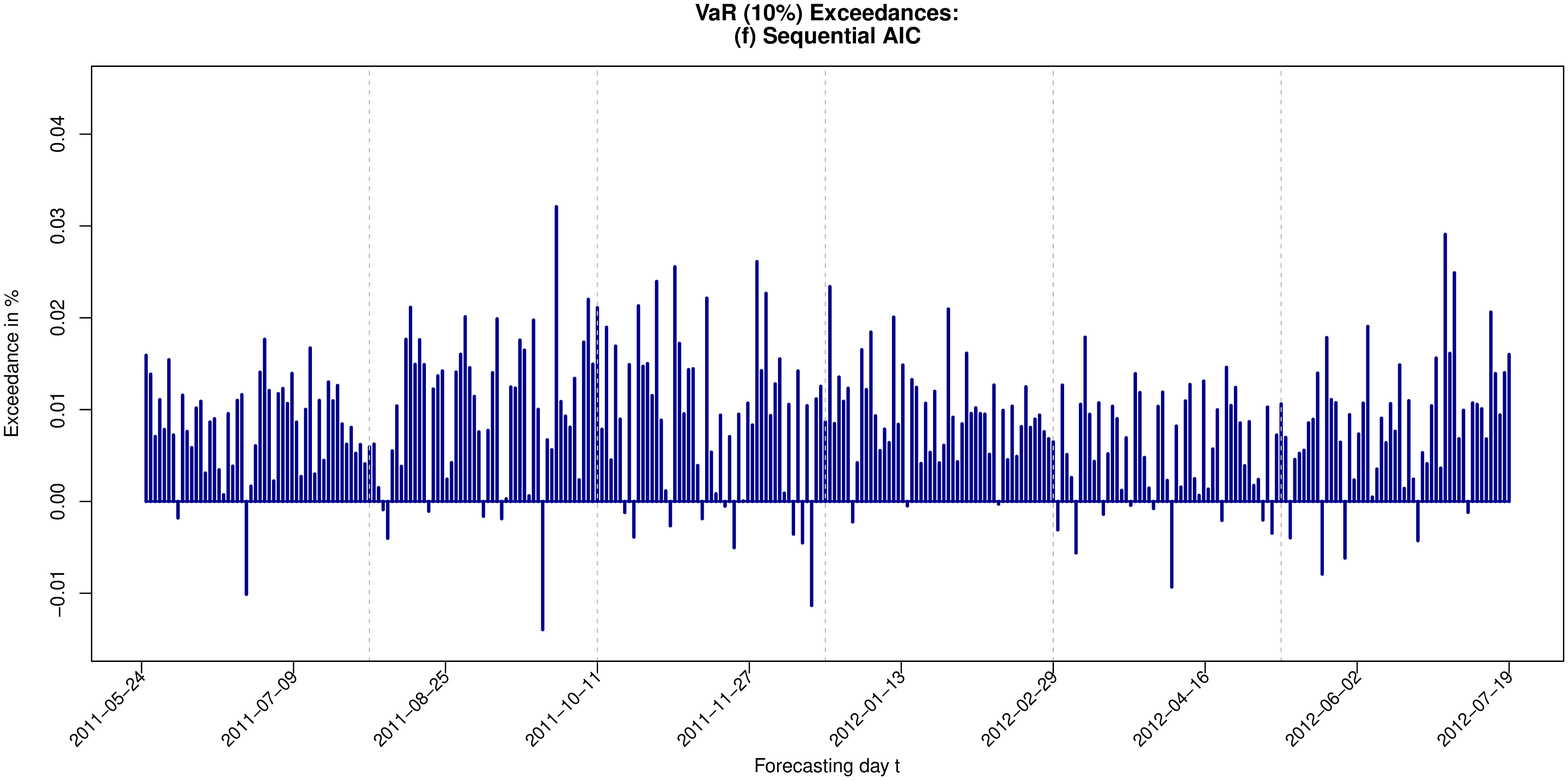}

\end{center}
\caption{{\bf Positive and negative VaR-exceedances for the Bernstein Vine and the parametric benchmark.} The figure shows plots of the positive and negative VaR-exceedances computed from the nonparametric Bernstein vine copula model (Panels (a), (c) and (e)) and the parametric benchmark vine model with the parametric pair-copulas chosen according to the sequential heuristic taken from the R-package \textit{CDVine} based on Akaike's Information Criterion (Panels (b), (d) and (f)). For both models, the $(1-\alpha)$-VaR is computed for confidence levels $\alpha\in\left\{2\%;5\%;10\%\right\}$. Both plots show results for the out-of-sample of $300$ trading days. The portfolio consists of the returns on the EURO STOXX 50 Price Index, US Treasury Bonds (30-year), France Benchmark 10-Year Government Bond Index, Gold Bullion LBM and Crude Oil-Brent one-month forward.}
\label{fig:exceed1}
\end{sidewaysfigure}

\newpage
\begin{sidewaysfigure}[htbp]

\begin{center}
\includegraphics[width=11.2cm]{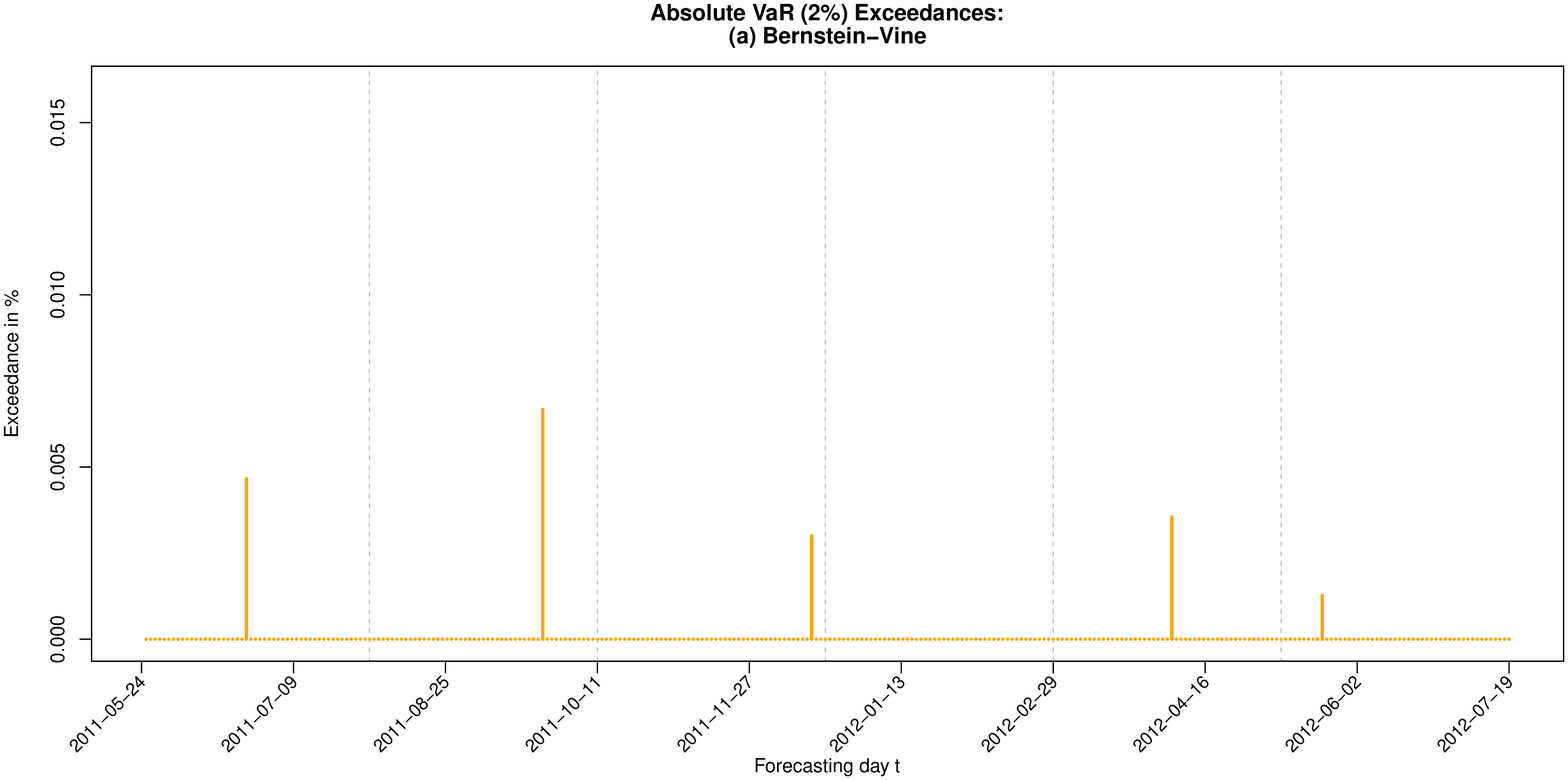}
\includegraphics[width=11.2cm]{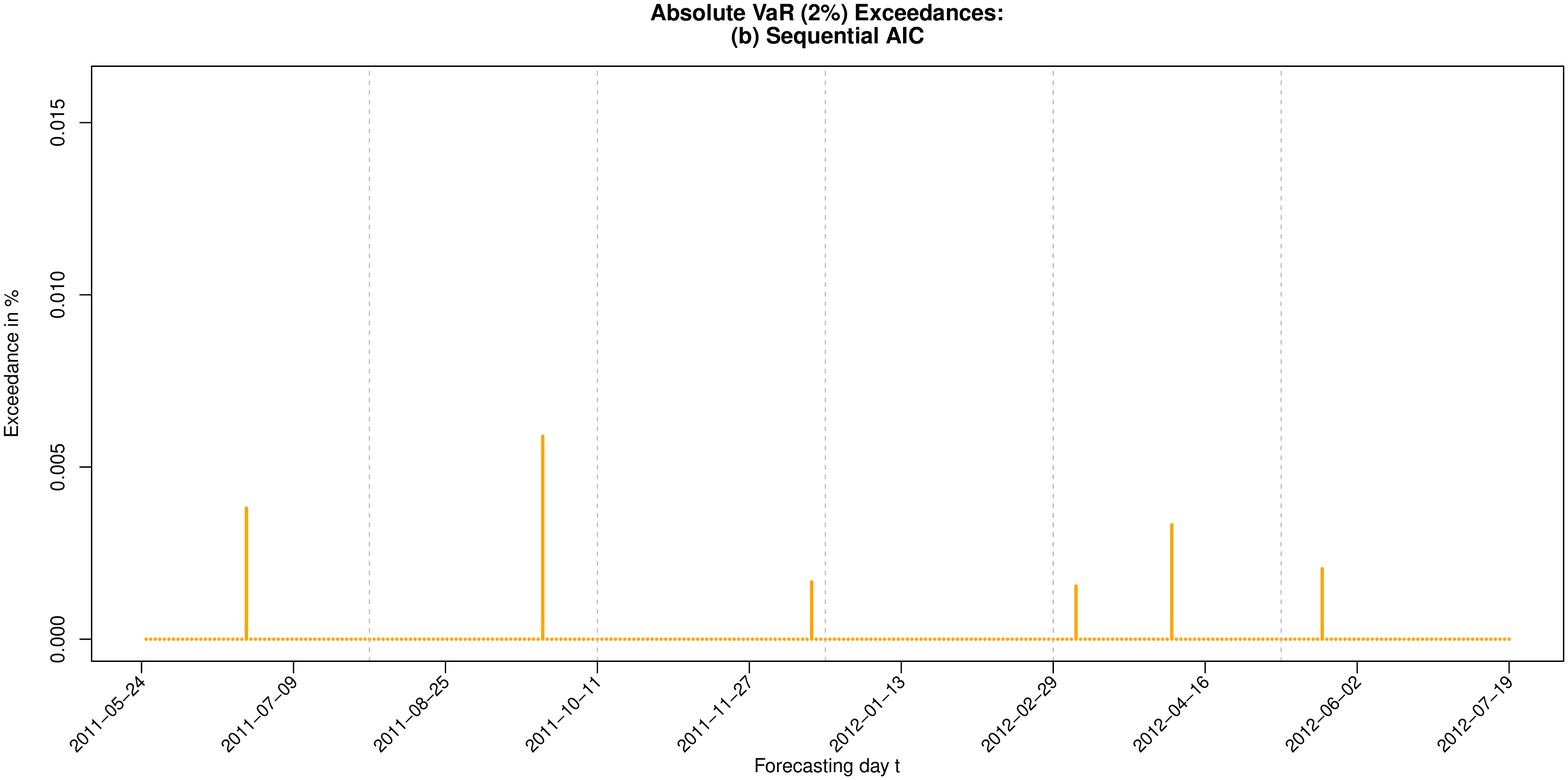}
\includegraphics[width=11.2cm]{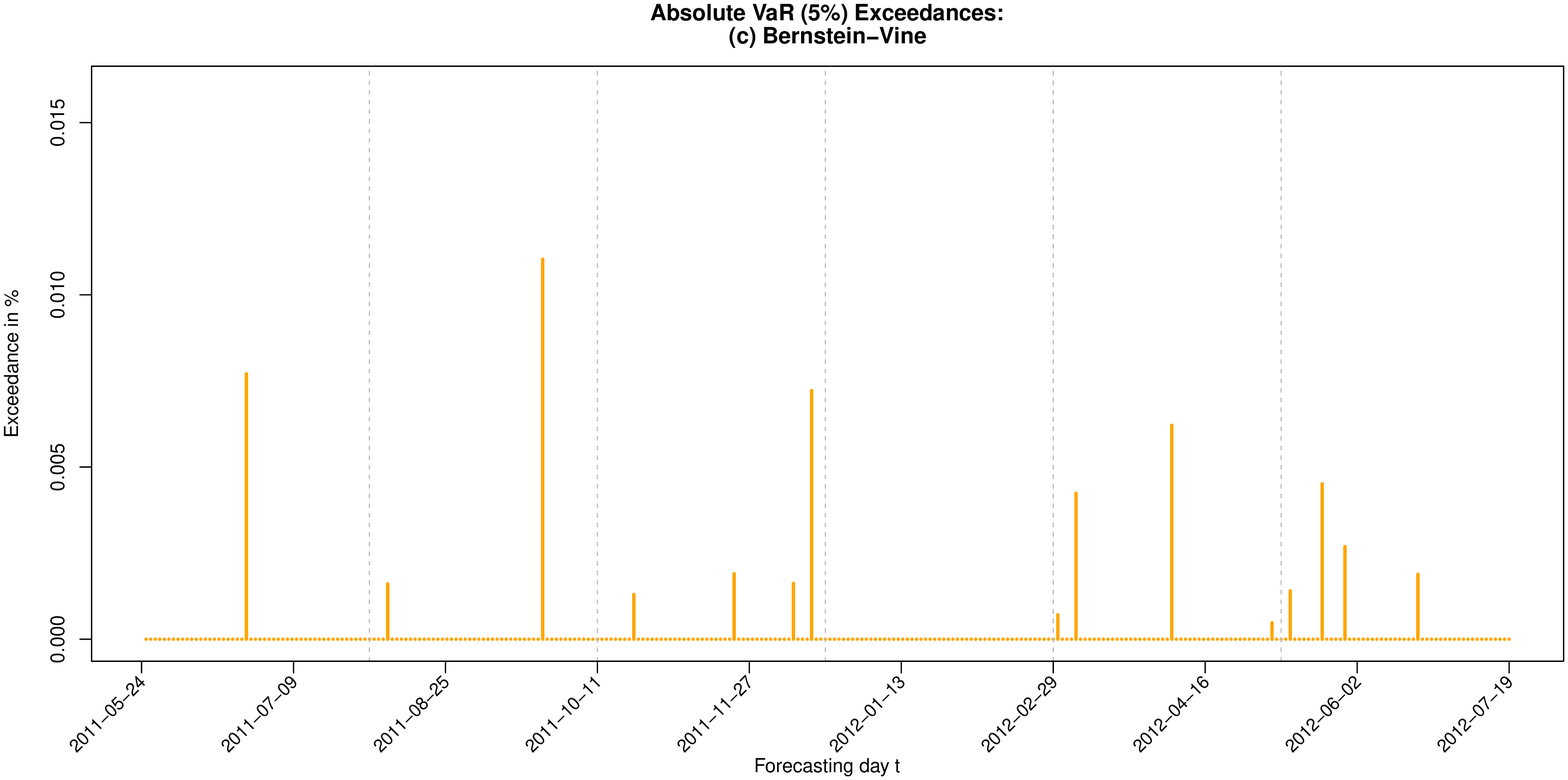}
\includegraphics[width=11.2cm]{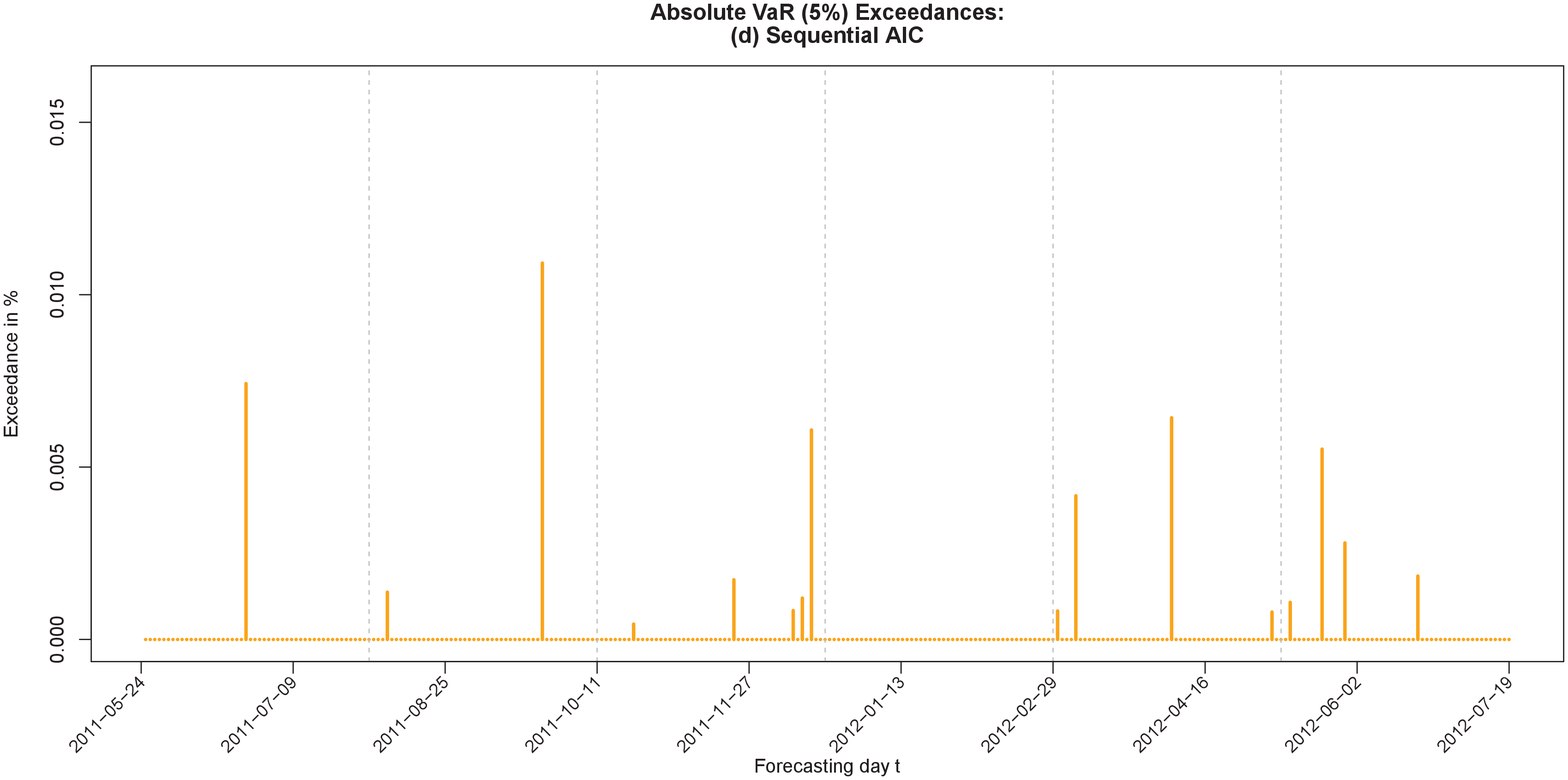}
\includegraphics[width=11.2cm]{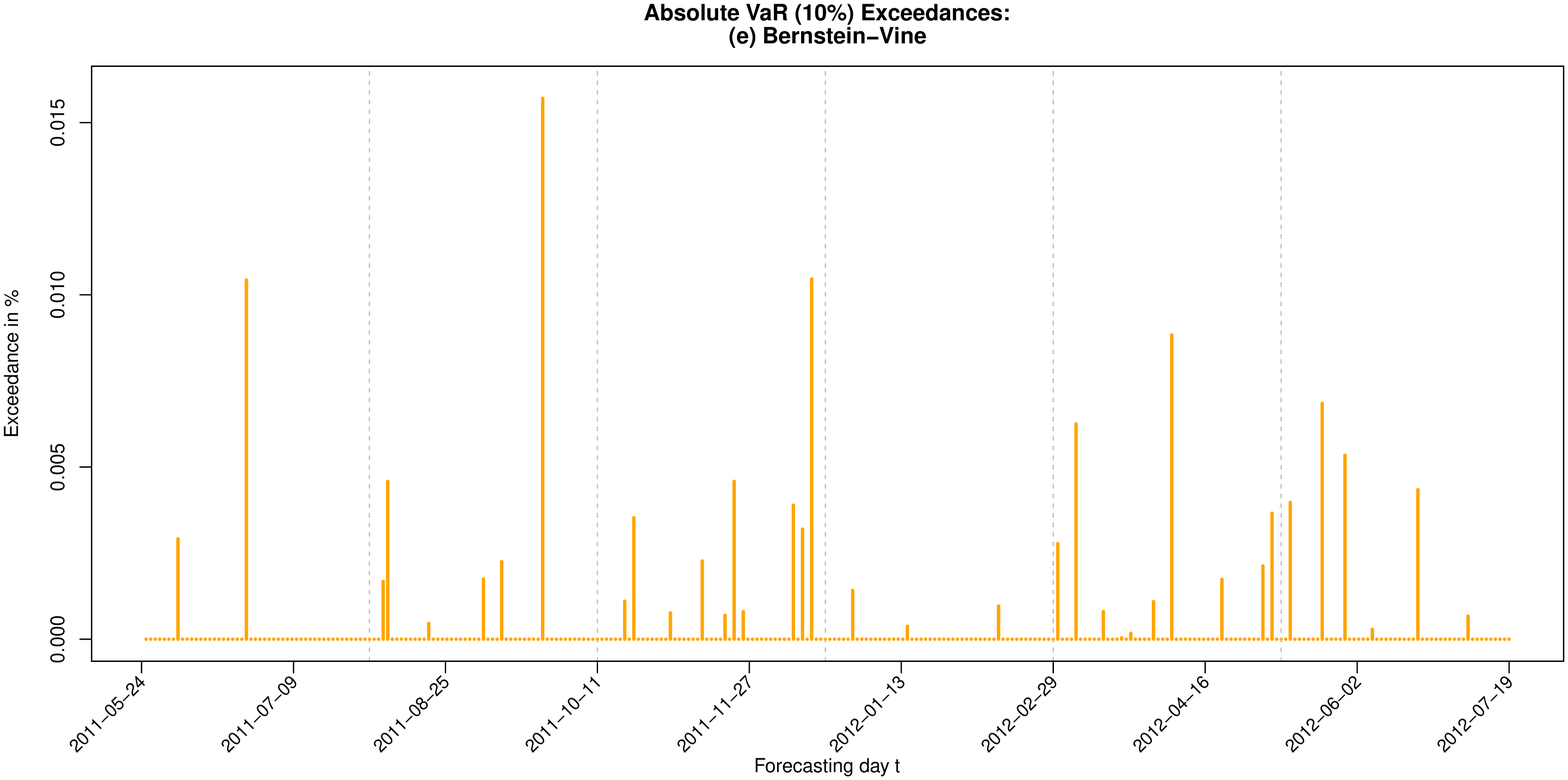}
\includegraphics[width=11.2cm]{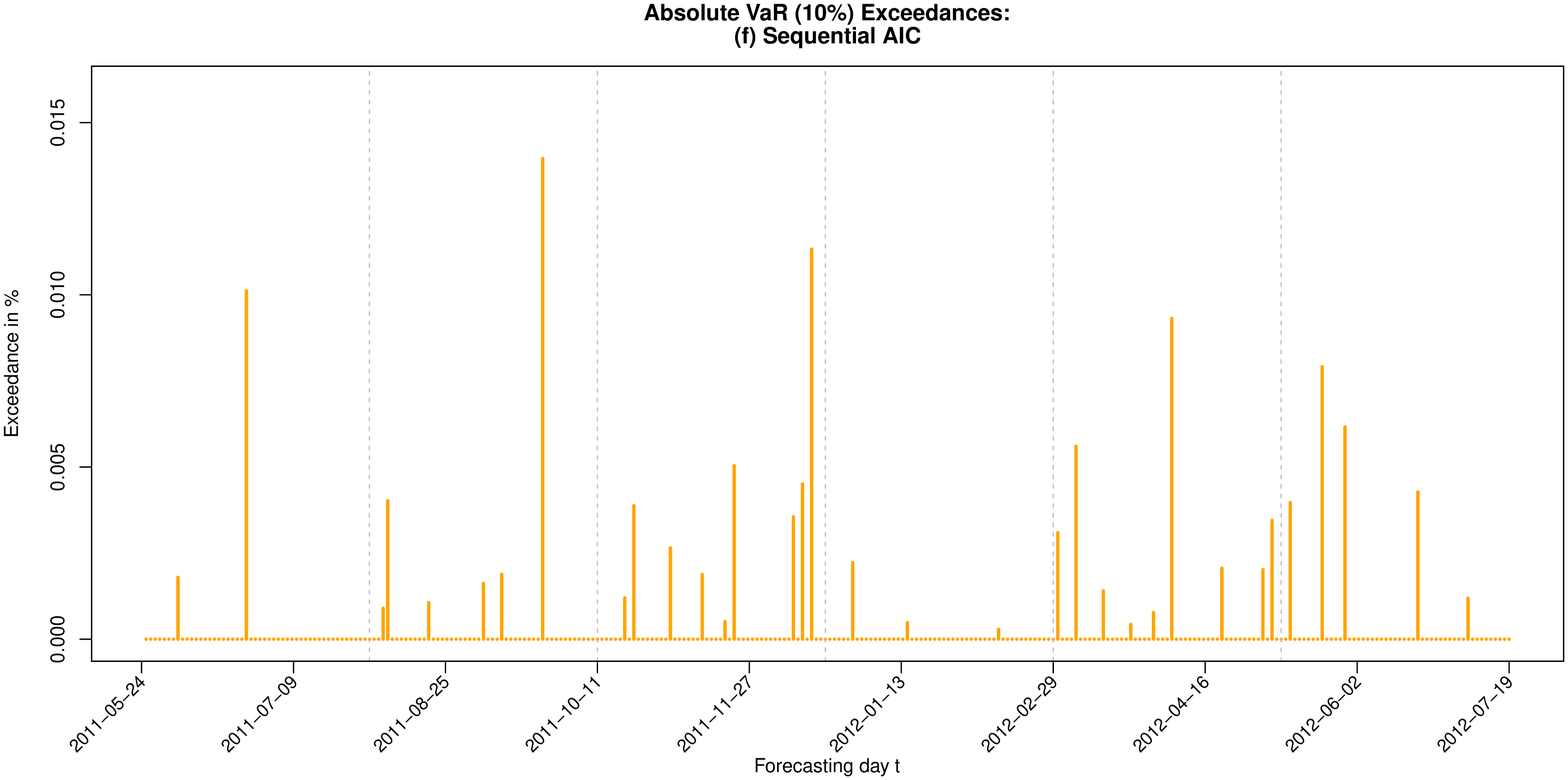}

\end{center}
\caption{{\bf Negative VaR-exceedances for the Bernstein Vine and the parametric benchmark.} The figure shows plots of the negative VaR-exceedances (i.e., losses below the daily VaR-forecasts) computed from the nonparametric Bernstein vine copula model (Panels (a), (c) and (e)) and the parametric benchmark vine model with the parametric pair-copulas chosen according to the sequential heuristic taken from the R-package \textit{CDVine} based on Akaike's Information Criterion (Panels (b), (d) and (f)). For both models, the $(1-\alpha)$-VaR is computed for confidence levels $\alpha\in\left\{2\%;5\%;10\%\right\}$. Both plots show results for the out-of-sample of $300$ trading days. The portfolio consists of the returns on the EURO STOXX 50 Price Index, US Treasury Bonds (30-year), France Benchmark 10-Year Government Bond Index, Gold Bullion LBM and Crude Oil-Brent one-month forward. For ease of presentation the losses exceeding the VaR-forecasts are shown as positive real numbers.}
\label{fig:exceed2}
\end{sidewaysfigure}

\begin{table}[ht]
\centering
\caption{{\bf Comparison of the Average Squared Error}. The table presents a comparison of the Average Squared Errors (ASE) of the parametric and nonparametric approximation to randomly specified vine copulas as well as the fraction of time (in \%) the sequential heuristic based on AIC breaks down due to either the numerical instability of the evaluation of the likelihood function and the parameter estimation or due to the ASE tending to infinity. The ASE is given in multiples of $10^{-3}$. All results are given in averages of $1,000$ simulations.}
\begin{tabular}{lcccc}
\toprule
& \multicolumn{2}{c}{\bf Sequential AIC} & \multicolumn{2}{c}{\bf Bernstein Pair-Copulas} \\
& {\bf ASE} & {\bf Instability (in \%)} & {\bf ASE} & {\bf Instability (in \%)}\\
\hline
\textit{Dimension $d=3$}\\
C-Vine ($n = 200$) & 0.040336 & 6.90 & 2.741199 & 0.00 \\
C-Vine ($n = 500$) & 0.030566 & 7.70 & 2.595884 & 0.00 \\
D-Vine ($n = 200$) & 0.044312 & 7.00 & 2.772305 & 0.00 \\
D-Vine ($n = 500$) & 0.023562 & 7.40 & 2.531201 & 0.00 \\
\hline
\textit{Dimension $d=5$}\\
C-Vine ($n = 200$) & 2.226120 & 17.30 & 5.063710 & 0.00 \\
C-Vine ($n = 500$) & 2.101471 & 17.80 & 4.864722 & 0.00 \\
D-Vine ($n = 200$) & 2.271577 & 18.30 & 5.103421 & 0.00 \\
D-Vine ($n = 500$) & 2.107821 & 18.90 & 4.912165 & 0.00 \\
\hline
\textit{Dimension $d=7$}\\
C-Vine ($n = 200$) & 4.128765 & 30.20 & 6.176894 & 0.00 \\
C-Vine ($n = 500$) & 3.843362 & 26.90 & 6.070333 & 0.00 \\
D-Vine ($n = 200$) & 4.021383 & 27.00 & 6.108655 & 0.00 \\
D-Vine ($n = 500$) & 3.880709 & 28.70 & 6.083041 & 0.00 \\
\hline
\textit{Dimension $d=9$}\\
C-Vine ($n = 200$) & 5.394219 & 37.40 & 6.816040 & 0.00 \\
C-Vine ($n = 500$) & 5.219997 & 33.00 & 6.782382 & 0.00 \\
D-Vine ($n = 200$) & 5.021444 & 36.40 & 6.741291 & 0.00 \\
D-Vine ($n = 500$) & 4.956846 & 37.10 & 6.736978 & 0.00 \\
\hline
\textit{Dimension $d=11$}\\
C-Vine ($n = 200$) & 6.184304 & 45.00 & 7.308551 & 0.00 \\
C-Vine ($n = 500$) & 5.972068 & 39.60 & 7.187704 & 0.00 \\
D-Vine ($n = 200$) & 5.843754 & 46.50 & 7.164167 & 0.00 \\
D-Vine ($n = 500$) & 5.688924 & 42.70 & 7.124501 & 0.00 \\
\hline
\textit{Dimension $d=13$}\\
C-Vine ($n = 200$) & 6.758818 & 50.00 & 7.544622 & 0.00 \\
C-Vine ($n = 500$) & 6.577154 & 46.60 & 7.533562 & 0.00 \\
D-Vine ($n = 200$) & 6.265431 & 54.50 & 7.421761 & 0.00 \\
D-Vine ($n = 500$) & 6.232211 & 50.60 & 7.404687 & 0.00 \\
\hline
\textit{Dimension $d=15$}\\
C-Vine ($n = 200$) & 7.159409 & 51.90 & 7.749707 & 0.00 \\
C-Vine ($n = 500$) & 7.112975 & 54.00 & 7.826334 & 0.00 \\
D-Vine ($n = 200$) & 6.680398 & 56.80 & 7.630245 & 0.00 \\
D-Vine ($n = 500$) & 6.598233 & 55.00 & 7.584712 & 0.00 \\
\bottomrule
\end{tabular}
\label{tab:simu}%
\end{table}

\newpage

\begin{sidewaystable}[ht]
\centering
\caption{{\bf Backtesting results}. The table presents the results of three different backtests performed on the out-of-sample forecasts for the portfolio VaR estimated from the vine copula models calibrated parametrically via the sequential selection of pair-copulas via AIC and the nonparametric modeling of the pair-copulas using Bernstein copulas, respectively. The three backtests are the test of conditional coverage, the unconditional and the conditional duration-based tests proposed in \citet{chris2}. The table reports the expected and the realized number of VaR-exceedances as well as the p-values for the three backtests. For both models, the backtesting results are reported for the $(1-\alpha)$-VaR for confidence levels $\alpha\in\left\{2\%;5\%;10\%;97.5\%\right\}$. For the $97.5\%$-VaR, exceedances are given under the assumption of a short position in the portfolio.}
\begin{tabular}{lccccc}
\toprule
& & \multicolumn{2}{c}{\bf Sequential AIC} & \multicolumn{2}{c}{\bf Bernstein Pair-Copulas} \\
& {\bf Exceedances} &  {\bf P-value} & {\bf Exceedances} &  {\bf P-value} & {\bf Exceedances} \\
& {\bf (expected)} &   & {\bf (realized)} &   & {\bf (realized)} \\

\hline
\textit{VaR $\alpha=2\%$}\\
Conditional Coverage & 6   & 0.994 & 6  & 0.850 & 5 \\
Unconditional Duration & 6 & 0.017 & 6  & 0.011 & 5 \\
Conditional Duration & 6   & 0.999 & 6  & 0.999 & 5 \\
\hline
\textit{VaR $\alpha=5\%$}\\
Conditional Coverage & 15  & 0.437 & 16  & 0.573 & 15 \\
Unconditional Duration & 15& 0.424 & 16  & 0.186 & 15 \\
Conditional Duration & 15  & 0.166 & 16  & 0.132 & 15 \\
\hline
\textit{VaR $\alpha=10\%$}\\
Conditional Coverage & 30  & 0.152 & 34  & 0.042 & 37 \\
Unconditional Duration & 30& 0.028 & 34  & 0.039 & 37 \\
Conditional Duration & 30  & 0.991 & 34  & 0.990 & 37 \\
\hline
\textit{VaR $\alpha=97.5\%$}\\
Conditional Coverage & 8  & 0.116 & 3  & 0.117 & 3 \\
Unconditional Duration & 8  & 0.053 & 3  & 0.052 & 3 \\
Conditional Duration & 8  & 1.000 & 3  & 1.000 & 3 \\
\bottomrule
\end{tabular}
\label{tab:backtest}%
\end{sidewaystable}

\end{document}